\documentclass[11pt]{article} 
\usepackage[left=2.5cm, right=2.5cm, top=1.785cm, bottom=2.0cm]{geometry}

\usepackage[utf8]{inputenc}
\usepackage{amsmath}
\usepackage{amsfonts}
\usepackage{amssymb}
\usepackage{mathtools}
\usepackage{varwidth}
\usepackage[linesnumbered, boxed, ruled]{algorithm2e}
\usepackage{listings}
\usepackage{mathrsfs}
\usepackage{times}
\usepackage{float}
\usepackage{color}
\usepackage{setspace}
\usepackage{color,soul}

\usepackage{amsmath,amsfonts,amsthm,eucal}
\usepackage{enumerate}
\usepackage{color}
\usepackage{tabularx}
\usepackage{siunitx}
\usepackage{empheq}
\usepackage[titletoc, title]{appendix}
\usepackage[tableposition=below]{caption}

\usepackage[
pdfstartview=XYZ,
bookmarks=true,
colorlinks=true,
linkcolor=blue,
urlcolor=blue,
citecolor=blue,
pdftex,
bookmarks=true,
linktocpage=true,   
hyperindex=true
]{hyperref}

\usepackage{lipsum}
\usepackage[FIGBOTCAP,TABTOPCAP,tight]{subfigure}

\usepackage[pdftex]{graphicx}
\usepackage{epstopdf}
\usepackage{booktabs} 
\usepackage{tikz,mathpazo}
\usetikzlibrary{shapes.geometric, arrows}
\usepackage{caption}

\theoremstyle{definition}

\usepackage{longtable}
\usepackage{adjustbox}
\usepackage{framed}
\usepackage{hyperref}

\usepackage[colorinlistoftodos]{todonotes}

\usepackage{amsmath}

\usepackage{nicefrac}
\usepackage{multirow}
\usepackage{hhline}
\usepackage{lineno}
\usepackage{import}
\usepackage{epstopdf}
\usepackage{subcaption}
\usepackage{mathtools}
\usepackage{transparent}

\usepackage{cancel}
\usepackage{empheq}
\usepackage{booktabs}
\usepackage{cleveref}

\crefname{equation}{Eq.}{Eqs.}
\usepackage{soul}
\usepackage{courier}

\SetKwInput{KwInput}{Input}
\SetKwInput{KwOutput}{Output}

\definecolor{tbf}{RGB}{255,0,0} 
\definecolor{txue}{RGB}{0,0,255}

\usepackage{authblk}
\usepackage{listings}
\usepackage{titlesec}
\usepackage[numbers,sort&compress]{natbib}

\newcommand\keywords[1]{\textbf{Keywords}: #1}

\title{\Large Solving contact problems using Fiber Monte Carlo} 

\begin{document}

\author[1]{\normalsize Xinyu Wang}
\author[1]{\normalsize Weipeng Xu}
\author[1]{\normalsize Tianju Xue
\footnote{\textit{cetxue@ust.hk}  (corresponding author)}}
\affil[1]{\footnotesize Department of Civil and Environmental Engineering, Hong Kong University of Science and Technology, HKSAR, China}
\date{}
\maketitle

\vspace{-30pt}

\begin{abstract}
Computational modeling of contact is fundamental to many engineering applications, yet accurately and efficiently solving complex contact problems remains challenging. 
In this work, we propose a new contact algorithm that computes contact forces by taking the gradient of an energy function of the contact volume (overlap) with respect to the geometry descriptors. 
While elegant in concept, evaluating this gradient is non-trivial due to the arbitrary geometry of the contact region. Inspired by the recently proposed Fiber Monte Carlo (FMC) method, we develop an algorithm that accurately computes contact forces based on the overlap volume between bodies with complex geometries. 
Our computational framework operates independently of mesh conformity, eliminating the need for master–slave identification and projection iterations, thus handling arbitrary discretizations. 
Moreover, by removing explicit complementarity constraints, the method retains a simple structure that can be easily incorporated into existing numerical solvers, such as the finite element method.
In this paper, numerical examples cover a wide range of contact scenarios, from classical small-deformation static contact to complex large-deformation dynamic contact in both two- and three-dimensional settings with nonlinear material behavior. 
These cases include Hertzian contact for small-deformation verification; contact between wedge- and cone-shaped bodies to assess pressure and displacement predictions at non-smooth boundaries; contact involving Neo-Hookean hyperelastic materials for evaluating nonlinear responses under finite deformation; and dynamic collision cases to examine transient behavior.
\end{abstract}


\section{Introduction}
\label{sec:intro}
The numerical treatment of contact mechanics using the finite element method (FEM) has been extensively studied for decades in computational mechanics, as contact plays a critical role in numerous real-world engineering applications; see the overview in~\cite{wriggers2006computational, wriggers1995finite}. However, the computational solution of contact problems is inherently challenging because the contact region is not known a priori, leading to nonlinear and non-smooth boundary value problems even in the absence of other nonlinearities (e.g., material and geometric nonlinearities)~\cite{de2017computational}. The numerical challenges in computational contact mechanics mainly involve two aspects: (1) the enforcement of contact constraints, and (2) the discretization of contact interfaces. 

A general numerical approach is to formulate the contact problem as a constrained optimization problem, whose theoretical foundations lie in variational inequalities, wherein the total potential energy of the system is minimized subject to the kinematic contact constraints~\cite{shen20222}. Two widely used schemes for enforcing contact constraints are the penalty method~\cite{sewerin2020finite, kang2023improved, peric1992computational} and the Lagrange multiplier method~\cite{weyler2012contact, papadopoulos1998lagrange, tur2009mortar}, both of which can be employed within the FEM framework. While each scheme transforms the constrained problem into an unconstrained one, they exhibit distinct intrinsic characteristics~\cite{simo1985perturbed}. The Lagrange multiplier method exactly enforces the impenetrability, introducing extra variables-Lagrange multipliers. However, this approach results in zero diagonal terms in the system matrix, leading to potential convergence issues in Newton-Raphson iterations~\cite{zienkiewicz2005finite, popp2009finite}. The penalty method is widely used for its simple formulation without any additional variables. However, it only enforces constraints approximately, resulting in penetrations that are highly sensitive to the choice of penalty parameter. A small value induces excessive penetration, whereas a large value induces ill-conditioning of the stiffness matrix~\cite{belytschko2014nonlinear}. To address these issues, the augmented Lagrange method~\cite{simo1992augmented, pietrzak1999large} combines both advantages at the cost of introducing iterative updates and parameter sensitivity. Besides, as an alternative for impenetrability enforcement, Nitsche's method~\cite{annavarapu2014nitsche, mlika2017unbiased} weakly enforces contact constraints through a mesh-dependent stabilization term, allowing the consistent derivation of contact stresses from the displacement field.

Another challenging issue is the discretization of the contact interfaces, which is typically categorized into node-to-node (NTN), node-to-surface (NTS), and surface-to-surface (STS) schemes. The NTN~\cite{francavilla1975note, hughes1976finite} contact discretization scheme requires conforming meshes to accurately transfer pressure and remains limited to infinitesimal sliding or perfectly matched interfaces due to geometric compatibility constraints. The latter two schemes are chosen to handle non-matching meshes at the contact interface. The NTS~\cite{zavarise2009node, sun2023novel, HUGHES1976249} method enforces contact by preventing slave node penetration into master segments. The NTS method, while suitable for large sliding and non-matching meshes, has an inherent master–slave asymmetry that leads to patch test failure. Although dual-pass approaches (Mortar)~\cite{puso2020dual} or penalty regularization~\cite{zavarise2009modified} can alleviate this issue, they may introduce mesh interlocking~\cite{puso2004mortar}. Furthermore, The low-order discretization leads to $C_0$ geometry, causing abrupt normal changes and non-physical force oscillations. To address these issues, recent works employ geometric smoothing techniques~\cite{neto2014applying, sun2023novel}, such as isogeometric analysis (NURBS) ~\cite{agrawal2025three, de2011large, de2012mortar} and the virtual element method (VEM)~\cite{aldakheel2020curvilinear, wriggers2019virtual}. The STS~\cite{zavarise1998segment, el2001stability} contact algorithm enforces constraints between corresponding segments on the two contacting surfaces. It offers better accuracy in force resolution and passes the patch test. However, this approach requires satisfaction of the inf-sup condition~\cite{bathe2001inf}, introduces significant computational cost, and may necessitate remeshing for large deformations due to mesh non-conformity. 

Conventional contact algorithms involve three key components: constraint enforcement, kinematic description, and spatial discretization. However, they face three main challenges: (1) Contact constraints derived from variational inequality are highly nonlinear and nonsmooth, often leading to convergence difficulties; (2) The reliance of kinematic descriptions on local geometric projections for defining contact distances and normals compromises solution existence and uniqueness near non-smooth boundaries; (3) Contact discretization methods are constrained by element order, whereby low-order formulations introduce spurious normal jumps and force oscillations.

\begin{figure}[H] \centering
    {\includegraphics[width=0.8\textwidth]{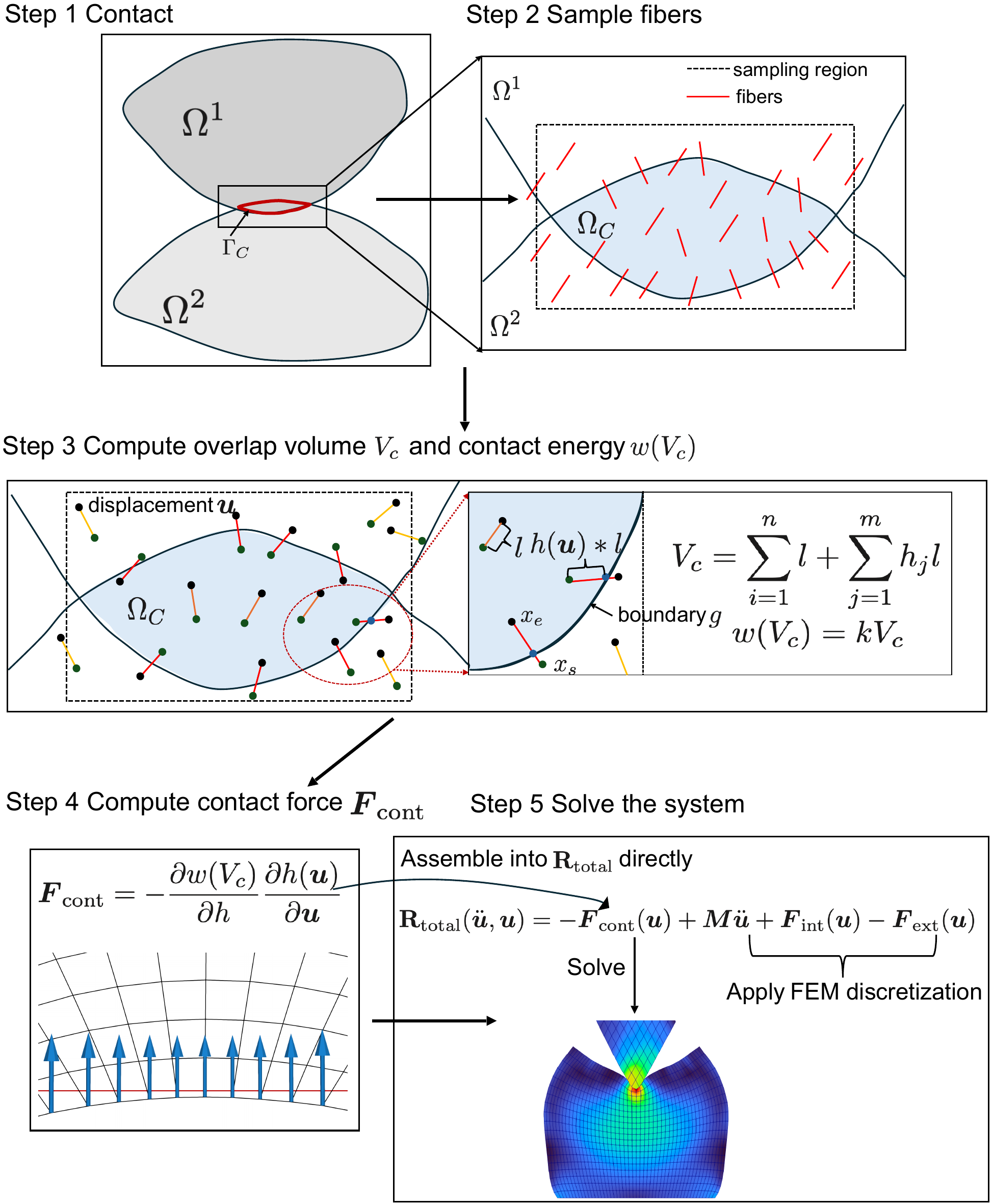}}
    \caption{A schematic flowchart of the proposed contact computational framework.} \label{fig:0.0}
\end{figure}

In this paper, we develop a fundamentally different computational contact method that aims to be more flexible, less dependent on geometry and meshes. Specifically, a contact energy model~\cite{feng2021energy}, defined in terms of the overlap volume between contact bodies, is constructed. The corresponding contact forces are then derived from the gradient of the energy function using the recently proposed Fiber Monte Carlo~\cite{richardson2024fiber} method. This method works by sampling line segments, enabling a differentiable volume estimation. In contrast, the traditional Monte Carlo method estimates volumes by sampling points and is not differentiable with respect to the geometry. Once computed, the contact forces are then directly assembled as additional nodal contributions into the global residual equation of FEM. The system is subsequently solved either explicitly in time using a symplectic integration scheme for dynamic problems or iteratively via a Newton-Raphson method for quasi-static simulations. The workflow implementing this computational framework is summarized in Fig.~\ref{fig:0.0}.  

Our approach is characterized by the following features:
\begin{itemize}
    \item Natural to handle complex geometries: the contact force and its normal direction are derived from the gradient of the contact energy function, overcoming the limitations of traditional projection methods near non-smooth boundaries, where issues of solution existence and uniqueness often arise (e.g., at concave corners or edges).
    \item Flexible with non-matching mesh topology: the method eliminates the need for contact pair search, master-slave identification, and projection iterations.  For instance, it robustly handles contact between coarse and fine meshes without additional mesh-conformity requirements.
    \item Simple to integrate into numerical solvers: contact forces are easily incorporated into discretized weak forms like an ``add-on'', ensuring consistent implementation in challenging contact scenarios, including complex kinematics (finite deformations and large sliding) as well as complex interface discretization (arbitrary non-matching meshes).
\end{itemize}

The work in this paper is organized as follows. Sec.~\ref{sec:computational framework} presents the mathematical description of the contact problem and detailed formulation of the energy-based contact model and the Fiber Monte Carlo method. Sec.~\ref{sec:problem} describes the contact boundary value problem in both strong and weak forms, reviews the spatial discretization schemes using the finite element method, and presents the solution procedure for the proposed method. Numerical examples are investigated in Sec.~\ref{Sec:examples} to illustrate the capability and performance of the proposed method, and finally some conclusions are drawn in Sec.~\ref{Sec:conclusion}.

Our code is available at \href{https://github.com/CMSL-HKUST/FMC_contact}{https://github.com/CMSL-HKUST/FMC$\_$contact}.

\section{Differentiable estimate of contact volume}
\label{sec:computational framework}
In this section, we introduce the methodology for modeling contact between two hyperelastic solids that undergo finite deformations in a dynamics setting. 
The problem is established within the continuum framework using a Lagrangian description. 
Then, we introduce a volume-based energy-conserving contact model, in which the contact energy is defined by the overlap volume between contacting bodies, and the contact force is obtained as the negative gradient of this contact energy. However, estimating the overlap volume for arbitrarily complex geometries and computing its gradient with respect to the geometry are non-trivial. To address this issue, we adopt the Fiber Monte Carlo method, which provides a differentiable estimate of the overlap volume via implicit differentiation.
\subsection{Problem description}
\label{sec:problem description}
\begin{figure}[H] \centering
{\includegraphics[width=0.8\textwidth]{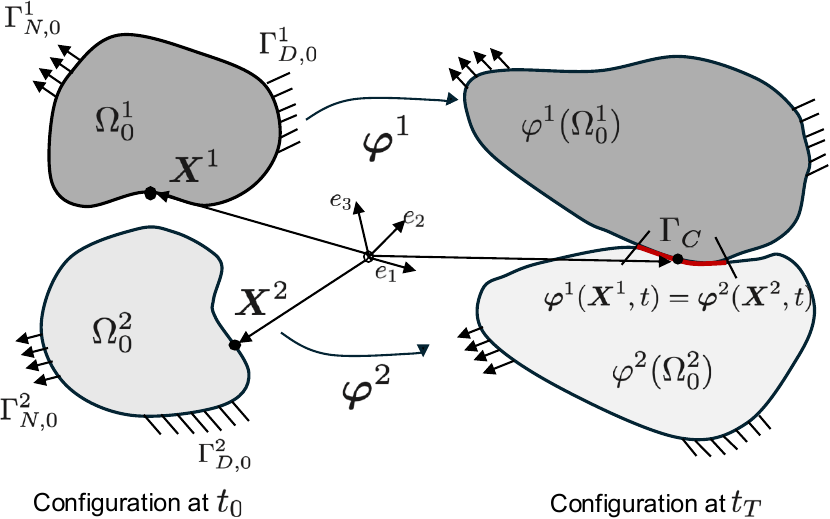}}
    \caption{A general finite-deformation contact problem.} \label{fig:1.0}
\end{figure}

We consider a general case of a system consisting of two deformable solids. In their reference (initial) configuration, they occupy regions $\Omega_0^i \subset \mathbb{R}^d$ (where $d$ denotes the spatial dimension, and $i = 1, 2$). Their motion and deformation lead to mutual contact constraints, as illustrated in Fig.~\ref{fig:1.0}. We denote by $\Gamma_0^i = \partial \Omega_0^i$ the boundary of the body in the reference configuration and by $\boldsymbol{N}^i$ the outward unit normal on $\Gamma_0^i$. The boundary $\Gamma_0^i$ is assumed to be composed of two disjoint parts $\Gamma_{D,0}^i$ and $\Gamma_{N,0}^i$ such that ${\Gamma_{D,0}^i \cup \Gamma_{N,0}^i = \Gamma_0^i}$ and ${\Gamma_{D,0}^i \cap \Gamma_{N,0}^i = \emptyset}$. We consider the dynamics of the system over a time interval $[0, T]$. For any time $t \in [0, T]$, the deformation of each body is described by a motion function 
$\boldsymbol{\varphi}^i: \Omega_0^i \times [0,T] \rightarrow \mathbb{R}^d$, which maps a material point $\boldsymbol{X}^i \in \Omega_0^i$ to its spatial counterpart $\boldsymbol{x}^i \in \Omega_t^i$, i.e., $\boldsymbol{x}^i = \boldsymbol{\varphi}^i(\boldsymbol{X}^i, t)$.
The corresponding displacement field for the body $i$ is defined as $\boldsymbol{u}^i(\boldsymbol{X}^i,t) = \boldsymbol{\varphi}^i(\boldsymbol{X}^i, t) - \boldsymbol{X}^i$,
and the velocity $\dot{\boldsymbol{u}}^i(\boldsymbol{X}^i,t) = \frac{\partial \boldsymbol{u}^i(\boldsymbol{X}^i, t)}{\partial t}$.
The total action functional of the system is given as:
\begin{align} \label{Eq:lagrangian}
S(\boldsymbol{u}^1, \boldsymbol{u}^2) = \int_{0}^{T} L(\boldsymbol{u}^1, \boldsymbol{u}^2, \dot{\boldsymbol{u}}^1, \dot{\boldsymbol{u}}^2)\,\mathrm{d}t,
\end{align}
under certain initial and boundary conditions. Here, we prescribe initial conditions for displacement $\boldsymbol{u}^i(\boldsymbol{X}^i, 0) = \boldsymbol{u}^i_0(\boldsymbol{X}^i) : \Omega_0^i \rightarrow \mathbb{R}^d$,
and velocity $\dot{\boldsymbol{u}}^i(\boldsymbol{X}^i, 0) = \boldsymbol{v}^i_0(\boldsymbol{X}^i) : \Omega_0^i \rightarrow \mathbb{R}^d, \quad \text{for } i=1,2$.
We also prescribe displacement $\boldsymbol{u}^i_D(\boldsymbol{X}^i, t) : \Gamma^i_{D,0} \times [0, T] \rightarrow \mathbb{R}^d$, and traction boundary conditions:
$\boldsymbol{t}^i_{N,0}(\boldsymbol{X}^i, t) : \Gamma^i_{N,0} \times [0, T] \rightarrow \mathbb{R}^d$, both of which can be time-dependent.

The classical contact mechanics problem is variationally formulated as a constrained minimization of the total potential energy, in which the satisfaction of unilateral inequality constraints is characterized by the Karush–Kuhn–Tucker (KKT) conditions~\cite{izmailov2003karush}. Within this framework, the Lagrangian of a system composed of two hyperelastic bodies, $\Omega_0^1$ and $\Omega_0^2$, which undergo finite deformations and are subject to mutual contact constraints (in the absence of body forces), is given by~\cite{marsden1994mathematical}:
\begin{align} \label{Eq:lagrangian_contact}
L(\boldsymbol{u}^1, \boldsymbol{u}^2, \dot{\boldsymbol{u}}^1, \dot{\boldsymbol{u}}^2)
&= \sum_{i=1}^{2} \Bigg[
\int_{\Omega_0^i} \frac{1}{2} \rho_0^i \, |\dot{\boldsymbol{u}}^i|^2  \mathrm{d}V
- \int_{\Omega_0^i} W^i(\boldsymbol{F}^i)  \mathrm{d}V
+ \int_{\partial \Omega_{N,0}^i} \boldsymbol{t}_{N,0}^i \cdot \boldsymbol{u}^i  \mathrm{d}A
\Bigg] \\
&\quad + \int_{\Gamma_C} \Phi(\boldsymbol{u}^1, \boldsymbol{u}^2)  \mathrm{d}\Gamma, \nonumber
\end{align}
where $\rho_0$ is the mass density in the reference configuration, 
$\boldsymbol{F} = \frac{\partial \boldsymbol{x}}{\partial \boldsymbol{X}} = \frac{\partial \boldsymbol{\varphi}(\boldsymbol{X},t)}{\partial \boldsymbol{X}}$
is the deformation gradient, and $W$ is the strain energy density function (per volume). The last term, $\Pi_c = \int_{\Gamma_C} \Phi(\boldsymbol{u}_1, \boldsymbol{u}_2) \mathrm{d}\Gamma$, represents the contact constraint contribution via KKT multipliers on the contact interface $\Gamma_C$: 
\begin{equation}
\Pi_c =  
\int_{\Gamma_C} \lambda_N g_N \mathrm{d}\Gamma,
\end{equation}
where $g_N$ denotes the normal gap function measuring the distance between surfaces of the two bodies in contact, and $\lambda_N$ represents the KKT multiplier interpreted as the normal contact pressure. Both variables satisfy the KKT conditions:
\begin{align}
g_N \geq 0,  \quad \lambda_N  \leq 0,  \quad g_N \, \lambda_N  = 0 \quad \text{on }\Gamma_C.
\end{align}

\subsection{Volume-based energy-conserving contact model}
\label{sec:energy model}

In contrast to traditional constraint-based contact approaches, we adopt a volume-based energy-conserving contact model inspired by Feng et al.~\cite{feng2021energy, FENG2021113454}.
If the two contacting objects are allowed to overlap, we can assume that the contact potential energy depends on the overlap volume of bodies with arbitrary shapes in 2D or 3D, and the energy function is defined as:
\begin{equation}\label{eq_energy_model}
w = w(V_c),
\end{equation}
where $V_c$ denotes the overlap volume between the contacting bodies.
Therefore, the contact term $\Pi_c$ can be approximated as
\begin{equation}
\Pi_c \sim w\big(V_c(\boldsymbol{u})\big) = w\Big(  \int_{\Omega^1(\boldsymbol{u}) \cap \Omega^2(\boldsymbol{u})} 
\, \mathrm{d}x \Big).
\end{equation}
The energy function $w(V_c)$ is defined by the following power law:
\begin{equation}\label{eq_energy_function}
w(V_c) = k V_c^m.
\end{equation}
Here $k$ is considered as a normal stiffness associated with the material properties of contact bodies. In this work, we adopt a linear contact energy function ($m=1$). Compared with conventional contact approaches that are formulated as variational inequalities to enforce contact constraints, this volume-based energy-conserving contact model ensures well-defined normal contact force $\boldsymbol{F}_c$ given by the negative gradient of the contact potential energy:
\begin{equation}\label{eq_force}
\boldsymbol{F}_c := -\nabla_{\boldsymbol{u}} w({V}_c) = -  km V_c^{m-1} \nabla_{\boldsymbol{u}} V_c.
\end{equation}  
This formulation provides a straightforward evaluation of the normal contact characteristics, including the magnitude, direction, and point of application of the normal contact force, while being directly applicable to arbitrary convex and concave geometries without requiring additional geometric assumptions or case-specific parameters. However, it should be noted that the contact force $\boldsymbol{F}_c$ depends on $\nabla_{\boldsymbol{u}} V_c$ by the chain rule. 
Achieving such a differentiable volume estimate arising from arbitrary contact geometries is a non-trivial task. In the following section, we will introduce the approach to estimate not only the contact volume but also the contact force.

\subsection{Fiber Monte Carlo method}
\label{sec:FMC}
In computational physics and geometry processing~\cite{mohamed2020monte}, target quantities $\mathcal{L}(\theta)$, e.g., the overlap volume or area between intersecting shapes, can often be formulated as expectations of piecewise or discontinuous functions over complex domains. Monte Carlo (MC) methods~\cite{mooney1997monte}, as a widely adopted technique, provide a general mechanism for estimating such expectations, particularly when closed-form solutions are unavailable.
 
\begin{figure}[H] \centering
    \subfigure[]
    {\includegraphics[width=0.35\textwidth]{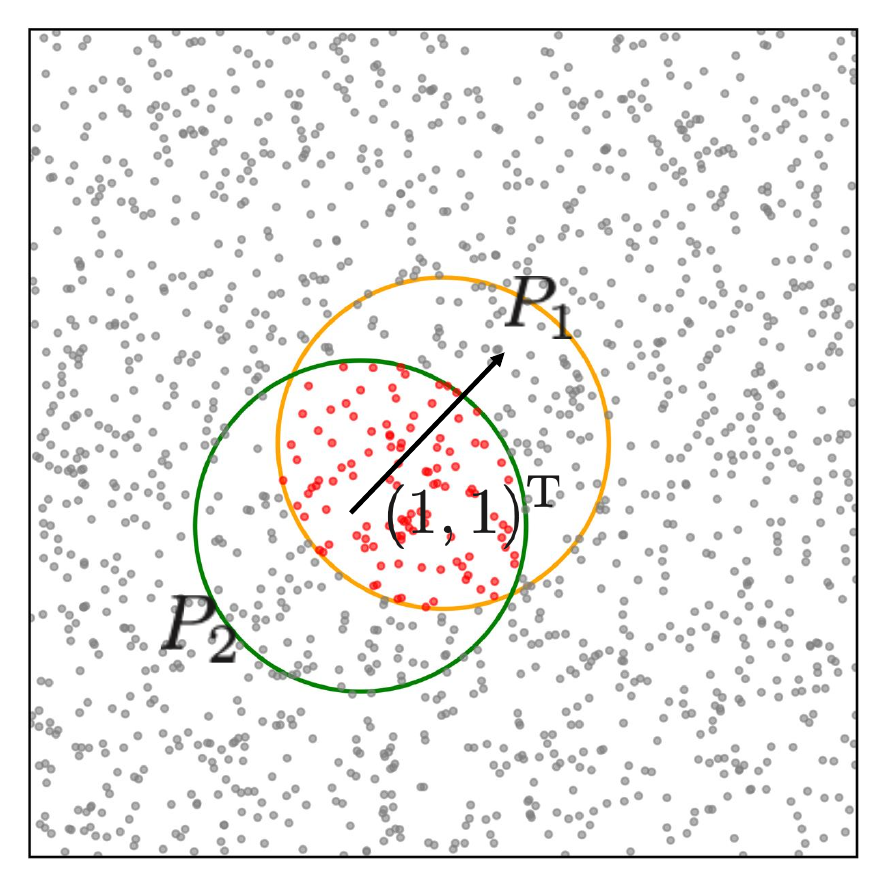}\label{fig:1.11.1}}
    \hspace{0.02\textwidth}
    \subfigure[]
    {\includegraphics[width=0.55\textwidth]{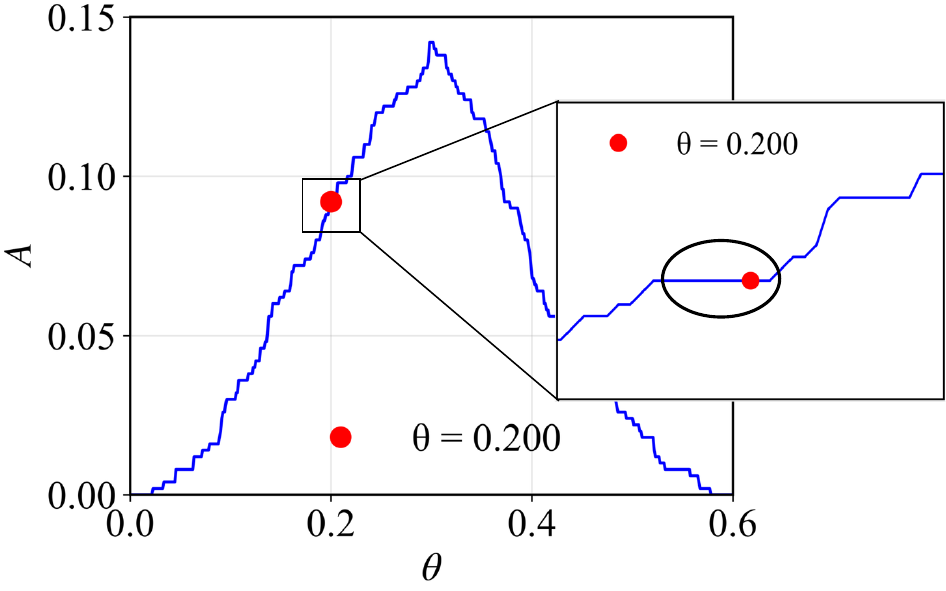}\label{fig:1.11.2}}
    \caption{Monte Carlo estimator: (a) the green, moveable shape $P_2$ is translated across the fixed orange shape $P_1$ along $(\theta, \theta)^\mathrm{T}$, and (b) evolution of the intersection area versus $\theta$.} \label{fig:1.11}
\end{figure}

To better illustrate the challenge of both the intersection measure and its gradient with respect to geometric parameters, we consider a simple case of two circles $P_1, P_2 \subset \mathbb{R}^2$, where $P_1$ is fixed and $P_2$ translates by an amount $\theta \in \mathbb{R}$ along the vector $(1,1)^\mathrm{T}$. The area of their intersection can be estimated by a Monte Carlo simulation with points uniformly sampled over the bounding domain $\Omega_s = [0, 1] \times [0, 1]$, as depicted in Fig.~\ref{fig:1.11}\subref{fig:1.11.1}. The intersection set is defined as $\Omega_P = \left\{ \boldsymbol{x} \in \mathbb{R}^2 \mid \boldsymbol{x} \in P_1 \cap P_2(\theta) \right\}$, and the target area is given by the integral:

\begin{equation}\label{eq:integral}
\mathcal{L}(\theta) = \int_{\Omega_s} \mathbb{I}_{\Omega_P}( \boldsymbol{x}) d\boldsymbol{x}.
\end{equation}
We define a random variable $\boldsymbol{X} \sim \mathrm{Uniform}(\Omega_s)$ with probability density $p(\boldsymbol{x}) = 1 / |\Omega_s| = 1$ on $\Omega_s$. The integral in Eq.~\eqref{eq:integral} can be equivalently expressed as the expectation of a random variable under uniform sampling over the domain:
\begin{equation}
\mathcal{L}(\theta) = |\Omega_s| \, \mathbb{E}_{\boldsymbol{X}}\left[\mathbb{I}_{\Omega_P}(\boldsymbol{X})\right] = \mathbb{E}_{\boldsymbol{X}}\left[\mathbb{I}_{\Omega_P}(\boldsymbol{X})\right]=\int_{\Omega_s} \mathbb{I}_{\Omega_P}(\boldsymbol{x}) \, p(\boldsymbol{x}) \, d\boldsymbol{x}.
\end{equation}
The indicator function $\mathbb{I}_{\Omega_P}(\cdot)$ is defined as:
\begin{equation}
\mathbb{I}_{\Omega_P}(\boldsymbol{x}) =
\begin{cases}
1, & \text{if } \boldsymbol{x} \in \Omega_P
 \\
0, & \text{otherwise}
\end{cases}.
\end{equation}
This expectation can be approximated using $N$ independent samples $\{\boldsymbol{x}_i\}_{i=1}^{N}$ drawn from the uniform distribution over $\Omega_s$, leading to the standard Monte Carlo estimator  $\hat{A}(\theta)$ for the true area  $A(\theta)$:
\begin{equation}
A(\theta) = \mathbb{E}_{\boldsymbol{X}}[I_{\Omega_P}(\boldsymbol{X})] \approx \frac{1}{N} \sum_{i=1}^{N} I_{\Omega_P}( \boldsymbol{x}_i ) = \hat{A}(\theta).
\end{equation}

The ultimate objective of this work is not only to obtain an estimate of the overlap area $A(\theta)$, but also an estimate of the derivative $\nabla_\theta A(\theta)$. However, while $\mathbb{E}_{\boldsymbol{X}}[I_{\Omega_P}(\boldsymbol{X})]$ is continuous and differentiable, the Monte Carlo estimator $\hat{A}(\theta)$ exhibits non-smooth, step-like behavior, as illustrated in Fig.~\ref{fig:1.11}\subref{fig:1.11.2}. 
The underlying reason for the non-differentiability of the standard Monte Carlo estimate can be clarified as follows:
\begin{equation}
\nabla_{\theta} \mathbb{E}_{\boldsymbol{X}} \left[ I_{\Omega_P}(\boldsymbol{X}) \right] \neq \mathbb{E}_{\boldsymbol{X}} \left[ \nabla_{\theta} I_{\Omega_P}(\boldsymbol{X}) \right].
\end{equation}
Specifically, the left-hand term $\nabla_{\theta} \mathbb{E}_{\boldsymbol{X}} \left[ I_{\Omega_P}(\boldsymbol{X}) \right]$  represents the desired derivative of the expectation:
\begin{equation}
\nabla_{\theta} \mathbb{E}_{\boldsymbol{X}} \left[ I_{\Omega_P}(\boldsymbol{X}) \right]=\nabla_{\theta} A(\theta) \neq 0.
\end{equation}
In contrast, the expectation of the gradient of the indicator function vanishes, i.e.,
\begin{equation}
\mathbb{E}_{\boldsymbol{X}}[{{\nabla_\theta}} I_{\Omega_P}(\boldsymbol{X})] = \mathbb{E}_{\boldsymbol{X}}[0] = 0.
\end{equation}
Therefore, the standard Monte Carlo approach using point samples can only estimate the contact volume but not the contact force, since the gradient of the estimator, i.e., $\nabla_{\theta} \hat{A}(\theta)$, vanishes and fails to provide a useful estimate.

\begin{figure}[H] \centering
    \subfigure[]
    {\includegraphics[width=0.35\textwidth]{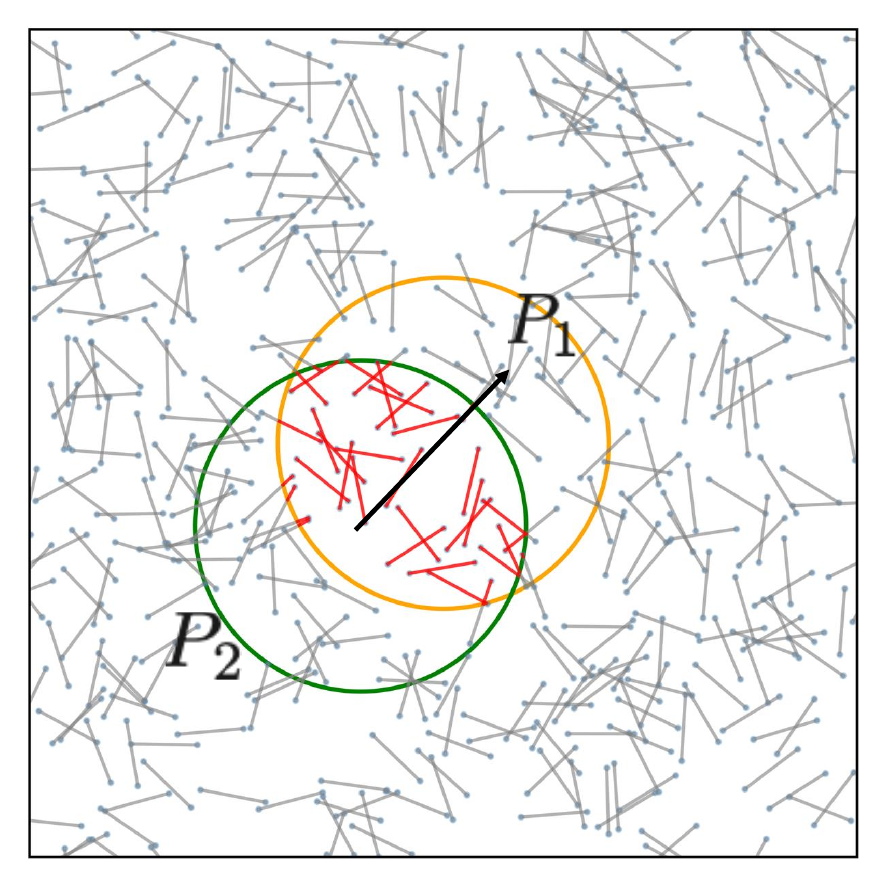}\label{fig:1.12.1}}
    \hspace{0.02\textwidth}
    \subfigure[]
    {\includegraphics[width=0.55\textwidth]{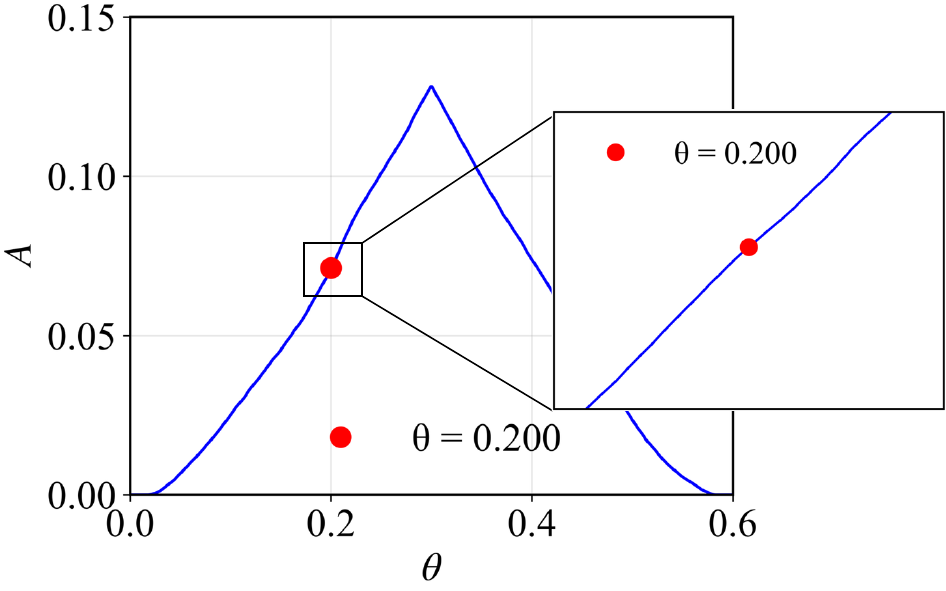}\label{fig:1.12.2}}
    \caption{Fiber Monte Carlo estimator: (a) the green, moveable shape $P_2$ is translated across the fixed orange shape $P_1$ along $(\theta, \theta)^\mathrm{T}$, and (b) evolution of the intersection area versus $\theta$.} \label{fig:1.12}
\end{figure}

To address the challenge of non-differentiable estimates and obtain correct gradients, we employ the Fiber Monte Carlo (FMC) method recently proposed by Richardson et al.~\cite{richardson2024fiber}.
FMC is a variant of standard Monte Carlo and it samples line segments (``fibers") instead of points to enable a differentiable estimator.
We utilize the same problem setup to demonstrate that differentiable estimation is achievable by FMC. As depicted in Fig.~\ref{fig:1.12}\subref{fig:1.12.1}, FMC samples fibers uniformly over the domain of integration rather than points. The detailed formulation of the sampling strategy employed in FMC is provided in Appendix~\ref{App:sample_method}. Fig.~\ref{fig:1.12}\subref{fig:1.12.2} shows that the FMC estimator produces a piecewise linear relationship between the estimated intersected area and the amount of translation $\theta$, which indicates a differentiable response. 

In FMC, the expectation can be decomposed using the law of total expectation:
\begin{equation}
\begin{aligned}
\mathbb{E}_{\boldsymbol{X}}[I_{\Omega_P}(\boldsymbol{X})] = \mathbb{E}_{f\sim F}\big[ \mathbb{E}_{\boldsymbol{X} | f} [I_{\Omega_P}(\boldsymbol{X}) ]\big],
\end{aligned}
\end{equation}
where $f$ denotes a sampled fiber, 
$\mathbb E_{f\sim F}$ denotes the expectation over the distribution of fibers, and $\mathbb E_{\boldsymbol{X}|f}[\cdot]$ denotes the expectation along points on each fiber. Under these definitions, expectation and differentiation can be interchanged:
\begin{equation}
\begin{aligned}
\nabla_{\theta}\mathbb{E}_{f}\Big[ \mathbb{E}_{\boldsymbol{X} \mid f} [I_{\Omega_P}(\boldsymbol{X}) ]\Big] = \mathbb{E}_{f}\Big[ \nabla_{\theta}\mathbb{E}_{\boldsymbol{X} \mid f} [I_{\Omega_P}(\boldsymbol{X}) ]\Big] 
\approx \frac{1}{N} \sum_{i=1}^{N} \nabla_{\theta}\mathbb{E}_{\boldsymbol{X} \mid f_i} \left[I_{\Omega_P}(\boldsymbol{X})\right],
\end{aligned}
\end{equation}
where $\nabla_{\theta}\mathbb{E}_{\boldsymbol{X} \mid f_i} \left[I_{\Omega_P}(\boldsymbol{X})\right]$ denotes the gradient of the conditional expectation for the $i$-th specific sampled fiber $f_i$, which can be exactly calculated using the implicit function theorem, as detailed in Appendix~\ref{App:Implicit function formulation}. Finally, the average of these gradients over all sampled fibers provides an unbiased estimator for the gradient of the expectation $\nabla_{\theta}\mathbb{E}_{\boldsymbol{X}} \left[I_{\Omega_P}(\boldsymbol{X})\right]$. 

\subsection{Verification of FMC method}
\label{sec:vetify}
Prior to discussing the computational framework in detail, we first assess the convergence and accuracy of the Fiber Monte Carlo (FMC) method for estimating both the overlap volume and its gradient, as introduced in Section~\ref{sec:FMC}. This verification is carried out using a set of two- and three-dimensional geometric benchmark cases.

\begin{figure}[H] \centering
    {\includegraphics[width=0.85\textwidth]{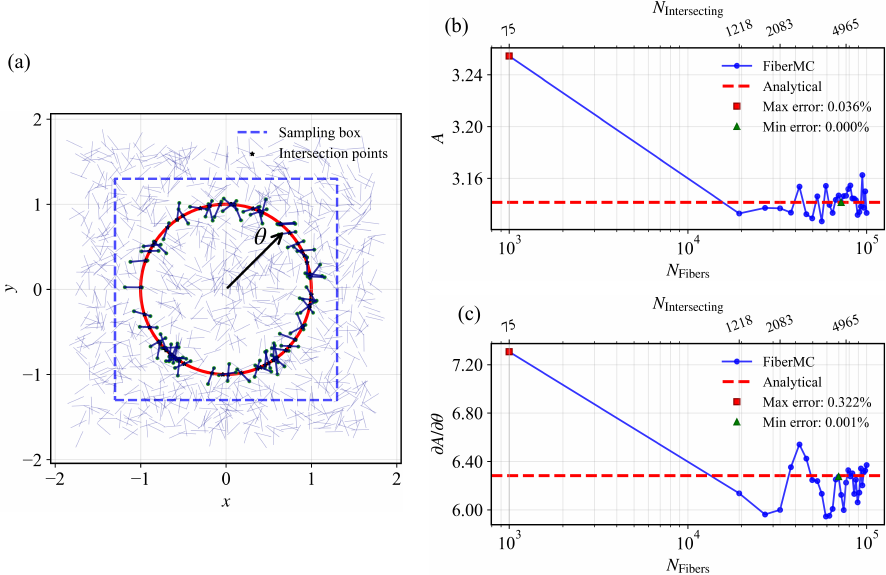}}
    \caption{Convergence analysis of the 2D test case with increasing number of intersecting fibers: (a) schematic of the sampling and geometry setup, (b) convergence of the estimated area, and (c) convergence of the derivative with respect to $\theta$.} \label{fig:1.2}
\end{figure}

\begin{figure}[H] \centering
    {\includegraphics[width=0.85\textwidth]{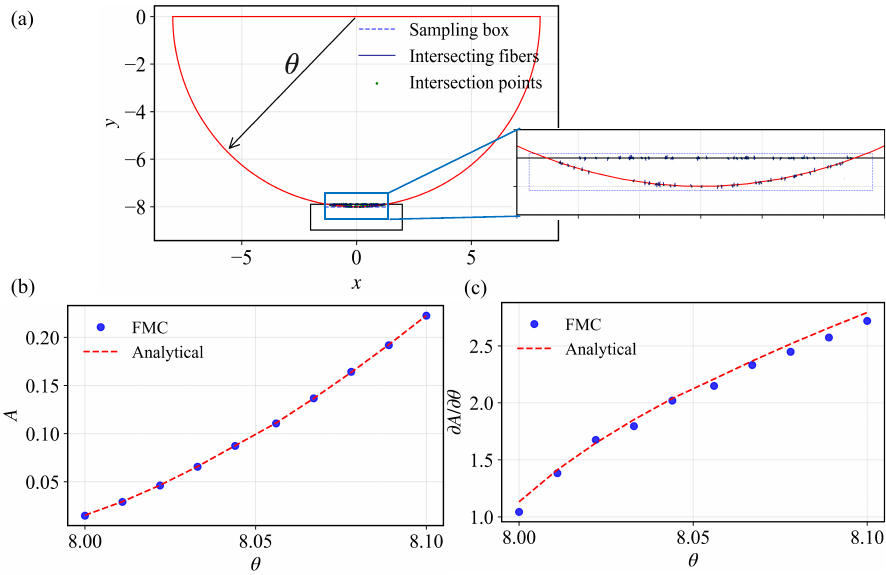}}
    \caption{Intersection of a semicircle and a block: (a) problem set up; (b) variation of the intersection area with the radius $\theta$; (c) first-order derivative of the intersection area, comparing the estimations obtained by FMC and analytical solutions. 
} \label{fig:1.31}
\end{figure}

\begin{figure}[H] \centering
    {\includegraphics[width=0.85\textwidth]{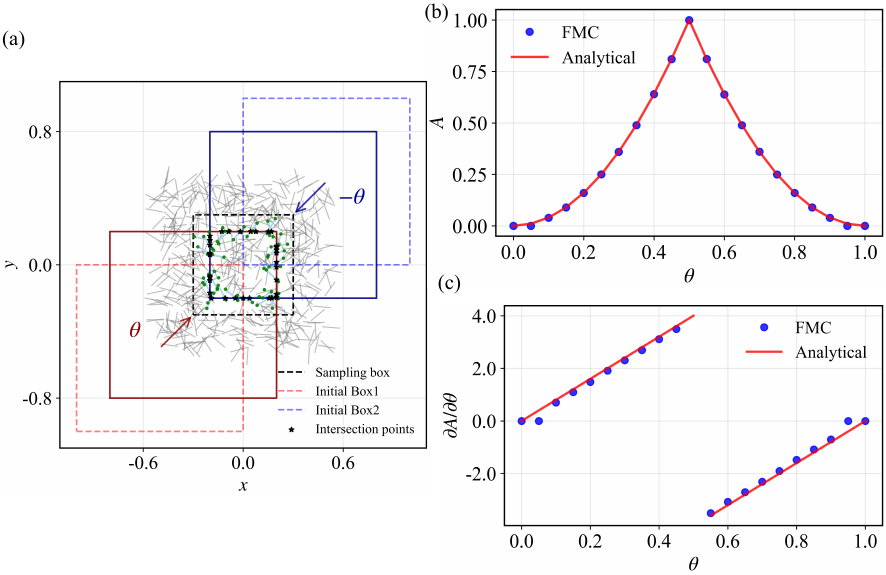}}
    \caption{Intersection of two translating squares: (a) problem set up; (b) variation of the intersection area with respect to $\theta$ as the two squares move toward each other along $(\pm\theta, \pm\theta)^\mathrm{T}$; (c) first-order derivative of the intersection area, comparing the estimations obtained by FMC and analytical solutions. 
} \label{fig:1.32}
\end{figure}

We first demonstrate the proposed FMC method in two dimensions by considering a circle of radius $\theta = 1$, as shown in Figure~\ref{fig:1.2}(a), and examining the convergence of both the area $A$ and its first-order derivative $\dfrac{\partial A}{\partial \theta}$ as the number of fibers increases.
The true area is $\pi \theta^2$, and the true derivative is $2\pi\theta$. Figure~\ref{fig:1.2}(b) and (c) show that the computed area and its derivative converge to these analytical values, with maximum relative errors of $0.036\%$ and $0.322\%$, respectively. To further assess the accuracy of the FMC estimates as the parameter $\theta$ varies, we evaluate the overlapping area $A$ between two objects and its derivative. Results in Figs.~\ref{fig:1.31} and~\ref{fig:1.32} demonstrate close agreement with the theoretical predictions over the considered range of $\theta$.

\begin{figure}[H] \centering
    {\includegraphics[width=0.8\textwidth]{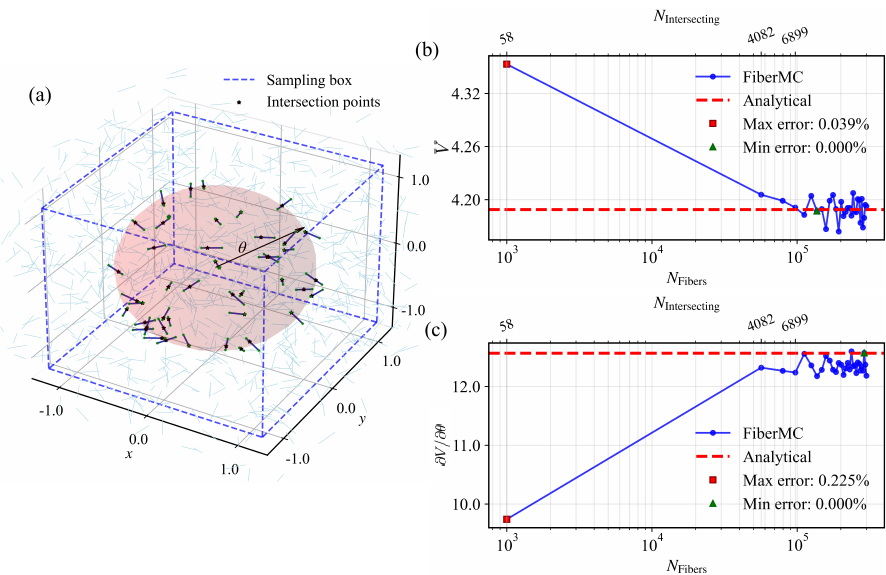}}
    \caption{Convergence analysis of the 3D test case with increasing number of intersecting fibers: (a) schematic of the sampling and geometry setup, (b) convergence of the estimated volume, and (c) convergence of the derivative with respect to $\theta$.} \label{fig:1.4}
\end{figure}

\begin{figure}[H] \centering
    {\includegraphics[width=0.8\textwidth]{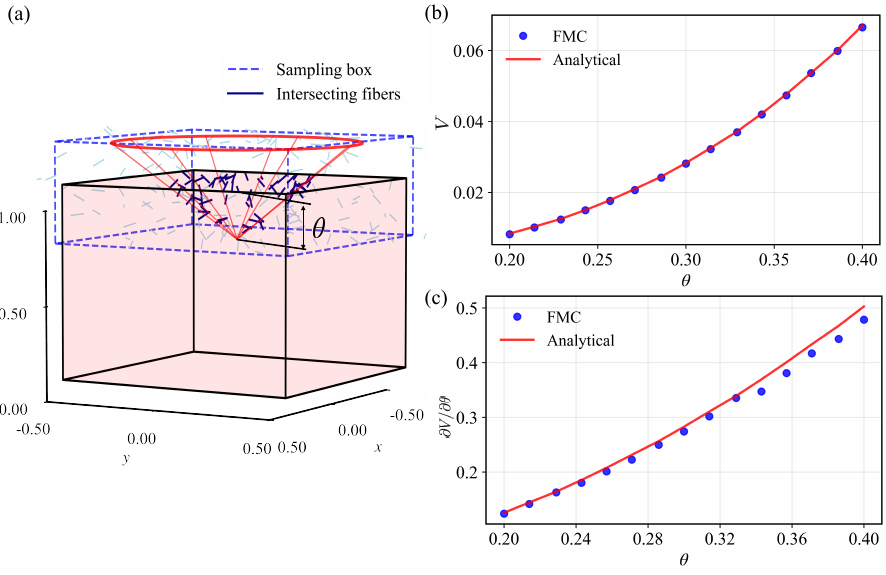}}
    \caption{Intersection of a fixed cube and a translating cone: (a) problem setup; (b) volume of intersection as a function of the translation amount $\theta$ along the $z$-axis; (c) derivative of the intersection volume with respect to $\theta$, comparing the estimations obtained by FMC and analytical solutions. 
} \label{fig:1.51}
\end{figure}

\begin{figure}[H] \centering
    {\includegraphics[width=0.8\textwidth]{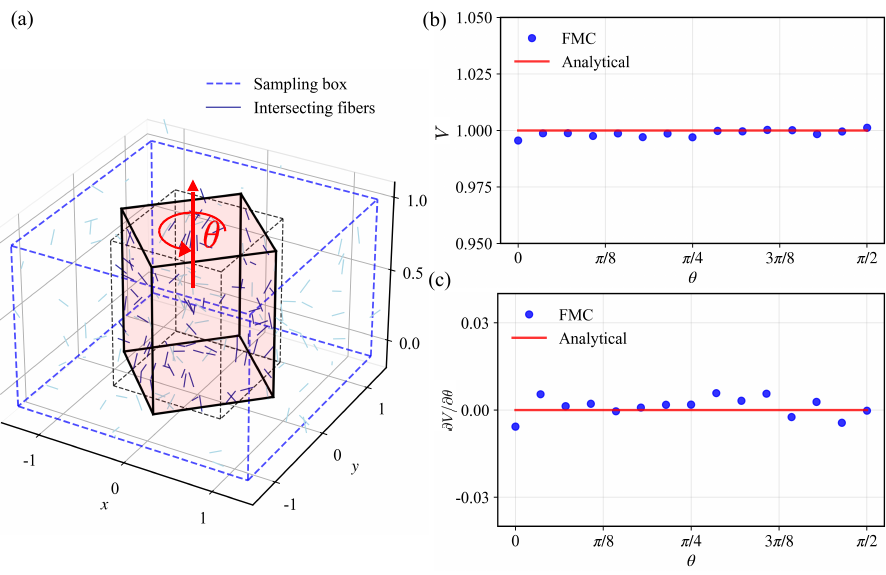}}
    \caption{Volume of the rotating cube: (a) problem setup; (b) volume of the cube as a function of the rotation angle $\theta$ about the $z$-axis; (c) derivative of the volume with respect to $\theta$, comparing the estimations obtained by FMC and analytical solutions.  
} \label{fig:1.52}
\end{figure}

To confirm the applicability of the proposed FMC method in three dimensions, we assess the convergence and accuracy of its estimates for volumes $V$ and the corresponding first-order derivatives with respect to the parameter $\theta$. First, we consider a sphere of radius $\theta = 1$, as depicted in Figure~\ref{fig:1.4}(a), and study the convergence of the estimated volume and its derivative with respect to $\theta$ as the number of sampled fibers increases. The exact solutions are $V = \tfrac{4}{3} \pi \theta^{3}$ and $\dfrac{\partial V}{\partial \theta} = 4 \pi \theta^{2}$. Figure~\ref{fig:1.4}(b) and (c) demonstrate that the computed overlapping volumes and their first-order derivatives converge to these analytical values, with maximum relative errors of $0.039\%$ for $V$ and $0.225\%$ for $\dfrac{\partial V}{\partial \theta}$. Furthermore, we study two three-dimensional overlap cases between objects, evaluating both the volume and its first-order derivative with respect to $\theta$ for each case. Results shown in Figs.~\ref{fig:1.51} and~\ref{fig:1.52} indicate close agreement with theoretical predictions.

The 2D and 3D benchmark cases demonstrate the convergence and accuracy of differentiable area/volume estimation obtained via the proposed FMC method. The results confirm that FMC provides accurate estimates for both contact volumes and their derivatives. Within the contact computational framework, these derivatives directly correspond to the normal contact forces. This verification paves the way for reliable contact force computation in nonlinear contact problems.

\section{Integration of contact algorithm into FEM} 
\label{sec:problem}

The proposed contact method based on energy models and Fiber Monte Carlo (as introduced in Section~\ref{sec:computational framework}) is straightforward to integrate into the finite element framework by simply adding the estimated contact force to the residual vector.
The strong form and the weak form of a standard finite-deformation elastic problem are first presented without contact. Subsequently, FE discretization is applied, resulting in a discrete system of equations, where contact forces are incorporated as additional nodal contributions to the global residual vector. Finally, the solution procedures for both static and dynamic contact analysis are described, respectively.
The contact algorithms within the finite element framework are implemented based on our open-source library JAX-FEM \cite{xue2023jax}.

\subsection{Strong form and weak form}
Consider a general boundary value problem involving two deformable solids in $\mathbb{R}^d$, denoted as $\Omega_0^i$ ($i=1,2$), which undergo finite deformations and experience mutual contact during their motion. The current position $\boldsymbol{x}^i$ of a material point initially located at $\boldsymbol{X}^i$ in body $\Omega_0^i$ at time $t$ is given by the displacement field $\boldsymbol{u}^i$ as: $\boldsymbol{x}^i = \boldsymbol{X}^i + \boldsymbol{u}^i(\boldsymbol{X}^i, t)$. The governing equations for elastodynamics at any point $\boldsymbol{X}^i$ of each solid $\Omega_0^i$ can be written in the reference configuration as:

\begin{subequations}\label{eq:system}
    \begin{align}
        \nabla_{\boldsymbol{X}} \cdot \boldsymbol{P}^i + \boldsymbol{b}_0^i &= \rho_0^i \partial_t^2 \boldsymbol{u}^i, \quad \forall \, \boldsymbol{X}^i \in \Omega_0^i \times [0,T], \label{eq:system1_ref} \\
        \boldsymbol{u}^i(\boldsymbol{X}^i, t) &= \overline{\boldsymbol{u}}^i, \quad \forall \, \boldsymbol{X}^i \in \Gamma_{D,0}^i \times [0,T], \label{eq:system2_ref}   \\ 
        \boldsymbol{P}^i \boldsymbol{N}^i &= \overline{\boldsymbol{t}}_0^i , \quad \forall \, \boldsymbol{X}^i \in \Gamma_{N,0}^i \times [0,T], \label{eq:system3_ref}  \\
        \boldsymbol{u}^i(\boldsymbol{X}^i, 0) &= \boldsymbol{u}^i_0(\boldsymbol{X}^i),  \quad \forall \, \boldsymbol{X}^i \in \Omega_0^i, \label{eq:system4_ref} \\
        \dot{\boldsymbol{u}}^i(\boldsymbol{X}^i, 0) &= \boldsymbol{v}^i_0(\boldsymbol{X}^i), \quad \forall \, \boldsymbol{X}^i \in \Omega_0^i, \label{eq:system5_ref} 
    \end{align}
\end{subequations}
where the divergence operator is defined in the reference configuration $\Omega_0^i$. The first Piola-Kirchhoff stress $\boldsymbol{P}$ is given by~${\boldsymbol{P} = \frac{\partial W}{\partial \boldsymbol{F}}}$ and $\boldsymbol{b}^i_0$ defines the body force. Eq.~\eqref{eq:system2_ref} and Eq.~\eqref{eq:system3_ref} represent the imposed Dirichlet and Neumann boundary conditions on $\Gamma_{D,0}^i$ and $\Gamma_{N,0}^i$, respectively, where the overbars in the notation (·) are used to denote prescribed boundary conditions of imposed displacements and surface tractions over the two boundaries. Here, $\Gamma_{D,0}^i \cap \Gamma_{N,0}^i = \emptyset$. The surface outward unit normal vector is denoted by $\boldsymbol{N}^i$.

To develop the weak form, we define the solution space $\mathcal{U}^i$ for the solid $\Omega_0^i$ as the set of all kinematically admissible deformations, and the weighting space $\mathcal{V}^i$ as the set of all admissible variations. These two functional spaces are then defined as:

\begin{subequations}
\begin{equation}
\mathcal{U}^i = \left\{ 
\boldsymbol{u}^i : \Omega_0^i \to \mathbb{R}^d \,\middle|\, 
\boldsymbol{u}^i \in [H^1(\Omega_0^i)]^d, \, 
\boldsymbol{u}^i = \bar{\boldsymbol{u}}^i \quad \forall \boldsymbol{X} \in \Gamma_{D,0}^i 
\right\},
\end{equation}
\begin{equation}
\mathcal{V}^i = \left\{ 
\delta \boldsymbol{u}^i : \Omega_0^i \to \mathbb{R}^d \,\middle|\, 
\delta \boldsymbol{u}^i \in [H^1(\Omega_0^i)]^d, \, 
\delta \boldsymbol{u}^i = \boldsymbol{0} \quad \forall \boldsymbol{X} \in \Gamma_{D,0}^i 
\right\},
\end{equation}
\end{subequations}
where $H^1(\Omega_0^i)$ denotes the Sobolev space~\cite{haslinger1981contact} of functions whose values and first derivatives are square-integrable over the reference domain $\Omega_0^i$. Accordingly, the weak form for dynamic equilibrium can be formulated as the problem of finding $\boldsymbol{u}^i \in \mathcal{U}^i$, $i=1,2$, such that:

\begin{equation}
\underbrace{\int_{\Omega_0^i} \rho_0^i \ddot{\boldsymbol{u}}^i \cdot \delta \boldsymbol{u}^i  dV}_{\delta \Pi_{\text{inertial}}}
+ \underbrace{\int_{\Omega_0^i} \boldsymbol{P}^i : \delta \boldsymbol{F}^i  dV}_{\delta \Pi_{\text{internal}}}
- \underbrace{\int_{\Omega_0^i} \boldsymbol{b}_0^i \cdot \delta \boldsymbol{u}^i  dV - \int_{\Gamma_{N,0}^i} \boldsymbol{t}_0^i \cdot \delta \boldsymbol{u}^i  dA}_{\delta \Pi_{\text{external}}}
= 0 \quad \forall \delta \boldsymbol{u}^i \in \mathcal{V}^i.
\label{eq:virtual_work}
\end{equation}
Here, the term $\delta \Pi_{\text{inertial}}$ represents the virtual work of the inertial forces, which is equivalent to the variation of kinetic energy of the system. The term $\delta \Pi_{\text{internal}}$ denotes the internal virtual work, which originates from the strain energy stored in the body. The term $\delta \Pi_{\text{external}}$ constitutes the external virtual work, contributed by the body forces $\boldsymbol{b}_0^i$ and the prescribed surface tractions $\boldsymbol{t}_0^i$.

\subsection{FEM discretization and contact enforcement}
\label{FEM discretization}
We begin the finite element discretization by constructing conforming finite-dimensional subspaces $\mathcal{U}_h^{i} \subset \mathcal{U}^i$ and 
$\mathcal{V}_h^{i} \subset \mathcal{V}^i$. The trial function 
$\boldsymbol{u}_h^{i} \in \mathcal{U}_h^{i}$ and test function 
$\delta \boldsymbol{u}_h^{i} \in \mathcal{V}_h^{i}$ are then introduced as the Galerkin approximations to their continuum counterparts $\boldsymbol{u}^i$ and $\delta \boldsymbol{u}^i$, respectively. These discretized fields and their gradients are expressed in terms of the basis functions as:
\begin{subequations}\label{eq:interpolation}
\begin{equation}
\boldsymbol{u}_h^{i}(\boldsymbol{X}) 
= \sum_{A=1}^{n^i} N_A(\boldsymbol{X}) \, \boldsymbol{d}_A^i, 
\quad 
\nabla_{\boldsymbol{X}} \boldsymbol{u}_h^{i}(\boldsymbol{X}) 
= \sum_{A=1}^{n^i} \boldsymbol{d}_A^i \otimes \nabla_{\boldsymbol{X}} N_A(\boldsymbol{X}),
\end{equation}
\begin{equation}
\delta \boldsymbol{u}_h^{i}(\boldsymbol{X}) 
= \sum_{A=1}^{n^i} N_A(\boldsymbol{X}) \, \delta \boldsymbol{d}_A^i, 
\quad 
\nabla_{\boldsymbol{X}} \delta \boldsymbol{u}_h^{i}(\boldsymbol{X}) 
= \sum_{A=1}^{n^i} \delta \boldsymbol{d}_A^i \otimes \nabla_{\boldsymbol{X}} N_A(\boldsymbol{X}),
\end{equation}
\end{subequations}
where $N_A(\boldsymbol{X})$ are the basis functions defined in the material configuration, $\boldsymbol{d}_A^i$ are the nodal displacement vectors of $\boldsymbol{u}_h^{i}$ for the body $i$,  $\delta \boldsymbol{d}_A^i$ are the corresponding nodal virtual displacement vectors, and $n^i$ is the total number of nodes for the body $i$.

Substituting the discretization of the trial and test functions ($\boldsymbol{u}_h^{i}$ and $\delta \boldsymbol{u}_h^{i}$) from Eq.~\eqref{eq:interpolation} into the weak form (Eq.~\eqref{eq:virtual_work}) yields its spatially discretized counterpart, which consists of the standard mechanical contributions: inertial, internal (elastic), and external terms. Having established the discretization of the standard mechanical terms, we next incorporate contact terms. In contrast to the standard mechanical terms, which are derived from the continuous weak form of the governing equations, contact interactions are implemented as discrete nodal forces directly incorporated into the global residual vector $\mathbf{R}^i_{\mathrm{total}}$ without additional interpolation or projection. Consequently, the dynamic equilibrium of each body $i$ is governed by the following discretized equation:

\begin{equation}
\mathbf{R}_{\mathrm{total}}^i(\ddot{\boldsymbol{d}}^i, \boldsymbol{d}^i) 
= \boldsymbol{M}^i \ddot{\boldsymbol{d}}^i + \boldsymbol{F}_{\mathrm{int}}^i(\boldsymbol{d}^i) - \boldsymbol{F}_{\mathrm{ext}}^i(\boldsymbol{d}^i) - \boldsymbol{F}_{\mathrm{cont}}^i(\boldsymbol{d}^i)
= \mathbf{0},
\quad \text{for } i = 1, 2
\label{eq:body_dynamic_residual}
\end{equation}
where $\boldsymbol{M}^i$ is the mass matrix. The global nodal displacement vector of body $i$ is defined as $\boldsymbol{d}^i = [\boldsymbol{d}^i_1, \boldsymbol{d}^i_2,...,\boldsymbol{d}^i_{n^i}]^T$, where $\boldsymbol{d}^i_{A}$ represents the displacement vector (a $d$-dimensional vector) at the $A$-th node of body $i$, and $\ddot{\boldsymbol{d}}^i$ denotes the corresponding global nodal acceleration vector. Similarly, $\boldsymbol{F}_{\text{int}}^i$ and $\boldsymbol{F}_{\text{ext}}^i$ denote the internal and external force vectors. The compact notation of Eq.~\eqref{eq:body_dynamic_residual} for a single body $i$ corresponds, at the nodal level, to contributions obtained by integrating over the reference domain $\Omega_0^i$ and its boundary $\Gamma_{N,0}^i$:
{\medmuskip=0.02\medmuskip
\begin{equation}
\mathbf{R}^i_{\text{total}} = 
\int_{\Omega_0^i}
\begin{bmatrix}
(\nabla_{\boldsymbol X} N_1)^T \boldsymbol{P}^i \\
(\nabla_{\boldsymbol X} N_2)^T \boldsymbol{P}^i \\
\vdots \\
(\nabla_{\boldsymbol X} N_{n^i})^T \boldsymbol{P}^i
\end{bmatrix} dV
-
\int_{\Omega_0^i}
\begin{bmatrix}
N_1 \boldsymbol{b}_0^i \\
N_2 \boldsymbol{b}_0^i \\
\vdots \\
N_{n^i} \boldsymbol{b}_0^i
\end{bmatrix} dV
-
\int_{\Gamma_{N,0}^i}
\begin{bmatrix}
N_1 \bar{\boldsymbol{t}}_0^i \\
N_2 \bar{\boldsymbol{t}}_0^i \\
\vdots \\
N_{n^i} \bar{\boldsymbol{t}}_0^i
\end{bmatrix} dA
+
\begin{bmatrix}
M_{L,1} \ddot{\boldsymbol{d}}^i_1 \\
M_{L,2} \ddot{\boldsymbol{d}}^i_2 \\
\vdots \\
M_{L,n^i} \ddot{\boldsymbol{d}}^i_{n^i}
\end{bmatrix}
-
\begin{bmatrix}
\boldsymbol F^i_{\text{cont},1} \\
\boldsymbol F^i_{\text{cont},2} \\
\vdots \\
\boldsymbol F^i_{\text{cont},n^i}
\end{bmatrix},
\end{equation}
}
where $M_{L,n^i}$ denotes the lumped mass at the $n$-th node, constructed by the row-sum technique applied to the consistent mass matrix $M^i$~\cite{taylor2013finite}.  This formulation completes the spatially discretized system, and the multi-body contact problem reduces to the simultaneous solution of the resulting coupled equations.

Following the volume-based energy model $w(V_c(\boldsymbol d))=kV_c$ in Sec.~\ref{sec:computational framework}, the normal contact force is defined as the negative gradient of $w(V_c(\boldsymbol d))$. The resulting contact force on body $i$, provided by our FMC-based differentiable estimation, is therefore given by:

\begin{equation}\label{eq:contactforce}
 \boldsymbol{F}^i_{\mathrm{cont}} = -\frac{\partial w(V_c(\boldsymbol d))}{\partial \boldsymbol{d}^i},
\end{equation}
where $\boldsymbol{d} = [\boldsymbol{d}^1, \boldsymbol{d}^2,...]^T$ is the global displacement vector of the contact system. The overlap volume $V_c$ is estimated using the FMC method:
\begin{equation}\label{eq:volume}
V_c = \sum_{i=1}^n l + \sum_{j=1}^m h_j(\boldsymbol d)l.
\end{equation}
The first term, $\sum_{i=1}^{n} l$, represents the total integral of fibers that are completely within the contact volume, where $l$ is the length of the fiber. The second term, $\sum_{j=1}^{m} h_j(\boldsymbol{d})l$, sums the contributions of all fibers that intersect the boundary within the contact volume between the bodies. $h_j(\boldsymbol{d})$ is an interpolation parameter defined (see Appendix~\ref{App:Implicit function formulation}) to locate the $j$-th intersection point $\boldsymbol{\alpha}$ where the fiber intersects the boundary. According to the FMC estimation of $V_c$ in Eq.~\eqref{eq:volume}, the contact force in Eq.~\eqref{eq:contactforce}. can be expressed as:

\begin{equation}
 \boldsymbol{F}^i_{\mathrm{cont}}
=-\frac{\partial w(V_c(\boldsymbol d))}{\partial h} \frac{\partial h(\boldsymbol{d})}{\partial \boldsymbol{d}^i},
\end{equation}
where the derivative of the contact potential energy $w(V_c)$ with respect to $h$ is straightforward. However, $h$ is implicitly defined by the deformed configuration as a function of the global displacement vector $\boldsymbol{d}$. Consequently, evaluating $\partial h / \partial \boldsymbol{d}$ requires to use the implicit function theorem. 
The detailed formulation is provided in the Appendix~\ref{App:Implicit function formulation}. For this purpose, a level set function representing the geometric boundary must be defined.
In this work, the geometric description uses the signed distance function (SDF). 
For any point $\boldsymbol{x} \in \mathbb{R}^d$, let $\text{dist}(\boldsymbol{x}, \partial\Omega) = \min_{\boldsymbol{y} \in \partial\Omega} \|\boldsymbol{x} - \boldsymbol{y}\|$ denote the Euclidean distance from $\boldsymbol{x}$ to the boundary $\partial\Omega$. Then, the SDF of a body $\Omega$ is defined as follows: 
\begin{equation}
    \text{SDF}(\boldsymbol{x}) = 
    \begin{cases} 
    -\text{dist}(\boldsymbol{x}, \partial \Omega), & \text{if } \boldsymbol{x} \in \Omega \text{ (inside)} \\
    \hphantom{-}0, & \text{if } \boldsymbol{x} \in \partial \Omega \text{ (on boundary)} \\
    \hphantom{-}\text{dist}(\boldsymbol{x}, \partial \Omega), & \text{if } \boldsymbol{x} \notin \Omega \text{ (outside)}.
    \end{cases}
\end{equation} 
For each body $i$, we denote its signed distance function by $\text{SDF}^i(\boldsymbol{x})$. 
Accordingly, the boundary of body $i$ is given by the zero level set of this function:
\begin{equation}
\partial \Omega^i = \{\, \boldsymbol{x} \in \mathbb{R}^d \mid \text{SDF}^i(\boldsymbol{x}) = 0 \,\}.
\end{equation}
For two bodies in mutual contact, the geometric description of the contact interface $\Gamma_C$ is given by:
\begin{equation}
\text{SDF} = \max(\text{SDF}^1, \text{SDF}^2),
\end{equation}
where $\text{SDF}^1$ and $\text{SDF}^2$ denote the signed distance functions associated with the two contacting bodies, respectively. 

\subsection{Solution procedures}
\label{sec:solution}
The spatially discretized equations in Eq.~\eqref{eq:body_dynamic_residual} are solved explicitly in time. 
We adopt a symplectic leapfrog scheme in its `kick–drift–kick' formulation~\cite{hairer2006structure}, which is widely used for Hamiltonian systems due to its long-term energy stability. 
The fully discretized update scheme is given by:
\begin{subequations}
\label{eq:leapfrog_kdk}
\begin{align}
    \boldsymbol{M}^i \boldsymbol{a}^i_n &= \boldsymbol{F}^i_{\mathrm{ext},n}  + \boldsymbol{F}^i_{\mathrm{cont},n} - \boldsymbol{F}^i_{\mathrm{int},n}
    \label{eq:acceleration} \\
    \boldsymbol{v}^i_{n+1/2} &= \boldsymbol{v}^i_n + \frac{1}{2} \boldsymbol{a}^i_n \Delta t
    \label{eq:kick1} \\
    \boldsymbol{u}^i_{n+1} &= \boldsymbol{u}^i_n + \boldsymbol{v}^i_{n+1/2} \Delta t
    \label{eq:drift} \\
    \boldsymbol{v}^i_{n+1} &= \boldsymbol{v}^i_{n+1/2} + \frac{1}{2} \boldsymbol{a}^i_{n+1} \Delta t.
    \label{eq:kick2}
\end{align}
\end{subequations}
A uniform time step $\Delta t$ is employed throughout all elastodynamics examples presented in this work, and damping effects are neglected. 
The overall solution procedure for explicit dynamic contact analysis comprising (i) computation of contact forces $\boldsymbol{F}^i_{\mathrm{cont}}$ within the proposed framework based on FMC, and (ii) time integration of the system state via the symplectic `kick--drift--kick' (KDK) scheme, is summarized in Algorithm~\ref{Alg:dynamics}. 
Within each time step, these two operations are performed sequentially, with the computed contact forces contributing as external nodal loads to the discretized momentum balance residual.

\begin{algorithm}[H]
\caption{Explicit Time Integration for Dynamic Contact}\label{Alg:dynamics}
\KwInput{Initialize displacement $\boldsymbol{u}^i_0$, velocities $\boldsymbol{v}^i_0$, accelerations $\boldsymbol{a}^i_0$, lumped-mass matrix $\boldsymbol{M}^i_{\text{LM}}$} 
\While{$t_n < T$}{
  \textbf{Kick 1:} Update $\boldsymbol{v}^i_{n+1/2}$ \tcp{See KDK equations in Eq.~\eqref{eq:kick1}}
  
  \textbf{Drift:} Update displacement $\boldsymbol{u}^i_{n+1}$  \tcp{See KDK equations in Eq.~\eqref{eq:drift}}
  
  Calculate contact forces $\boldsymbol{F}^i_{\text{cont}}$: \;
  \Indentp{1em}
  Randomly generate fibers within $\Omega_t$ for FMC sampling \tcp{See Appendix~\ref{App:sample_method}} 
  $\boldsymbol{F}^i_{\text{cont}} \leftarrow \text{ImplicitFunctionTheorem}(\boldsymbol{u}^i)$ \tcp{See Eq.~\eqref{eq:interpolant} to Eq.~\eqref{eq:implicit}}
  \Indentp{-1em}
   Calculate residual $\boldsymbol{F}^i_{\text{int}}$, $\boldsymbol{F}^i_{\text{ext}} \leftarrow \text{FEMdiscretization}(\boldsymbol{u}^i, \delta\boldsymbol{u}^i)$ \tcp{See Eq.~\eqref{eq:body_dynamic_residual}} 
  
  Solve $\boldsymbol{a}^i_{n+1} \leftarrow \text{Residual}(\mathbf{R}^i_{\text{total}}=0)$ \tcp{See Eq.~\eqref{eq:acceleration}} 
  \textbf{Kick 2:} Update $\boldsymbol{v}^i_{n+1}$ \tcp{See Eq.~\eqref{eq:kick2}} 
  $t_n \leftarrow t_{n+1}$ 
}
\KwOutput{Nodal states $\boldsymbol{u}^i_T$, $\boldsymbol{v}^i_T$}
\end{algorithm}

While the dynamics scheme described above is specifically designed for contact elastodynamics problems, its underlying computational framework can also be applied to quasi-static processes. By neglecting inertial effects, the method can be adapted to solve elastostatics problems, including quasi-static contact scenarios. This formulation leads to the following spatially discretized residual form for body $i$, obtained by reducing Eq.~\eqref{eq:body_dynamic_residual} to static equilibrium:

\begin{equation}
\mathbf{R}^i_{\text{total}}(\boldsymbol{u}^i) = \underbrace{\boldsymbol{F}^i_{\text{int}}(\boldsymbol{u}^i) - \boldsymbol{F}^i_{\text{ext}}(\boldsymbol{u}^i)}_{\mathbf{R}^i_{\text{mech}}(\boldsymbol{u}^{i})} - \boldsymbol{F}^i_{\text{cont}}(\boldsymbol{u}^i) = \mathbf{0}.
\label{eq:static_residual}
\end{equation}
This nonlinear system is solved using a Newton-Raphson scheme within a fixed-point iteration framework. The solver employs a two-loop iterative strategy. The outer fixed-point iteration resolves the contact-mechanics coupling by formulating a nonlinear equation:
\begin{equation}
\mathbf{R}^i_{\text{mech}}(\boldsymbol{u}^{i}_{n+1}) - \boldsymbol{F}^i_{\text{cont}}(\boldsymbol{u}^{i}_n) = 0, \quad \text{for } n = 1, 2, \ldots, N
\end{equation}
where $\boldsymbol{F}^i_{\text{cont}}$ is evaluated explicitly from the previous iteration. This equation is solved at each iteration by Newton's method, which iteratively linearizes the system and updates the displacement $\boldsymbol{u}^i_{n+1}$:
\begin{subequations}
\begin{align}
\left. \frac{\partial \mathbf{R}^i_{\text{mech}}}{\partial \boldsymbol{u}^i_{n+1}} \right|_{\boldsymbol{u}^{(i,m)}_{n+1}} \delta \boldsymbol{u}^{(i,m)} &= -\left[ \mathbf{R}^i_{\text{mech}}\left(\boldsymbol{u}^{(i,m)}_{n+1}\right) - \boldsymbol{F}^i_{\text{cont}}(\boldsymbol{u}^i_n) \right], \quad \text{for } m = 1, 2, \ldots, M  \label{eq:newton_iteration}   \\
\boldsymbol{u}^{(i,m+1)}_{n+1} &= \boldsymbol{u}^{(i,m)}_{n+1} + \delta \boldsymbol{u}^{(i,m)}.  \label{eq:update}
\end{align}
\end{subequations}
The finite element discretization and contact enforcement in this procedure remain consistent with those used in dynamic analyses. The solution procedure for quasi-static contact problem is detailed in Algorithm~\ref{Alg:static}. 

\begin{algorithm}[H]
\caption{Newton Solver for Quasi-Static Contact}\label{Alg:static}
\KwInput{Initial displacement $\boldsymbol{u}^i_{0}$} 
$n \leftarrow 0$ \\
$\boldsymbol{u}^i \leftarrow \boldsymbol{u}^i_{0}$ \tcp{Initialize current solution}
\While{$\|\boldsymbol{u}^i - \boldsymbol{u}^i_{n}\| > \epsilon_{\mathrm{outer}}$}{
  $\boldsymbol{u}^i_{n} \leftarrow \boldsymbol{u}^i$ \tcp{Store current solution for contact evaluation}
  Calculate contact force $\boldsymbol{F}^i_{\text{cont}} \leftarrow \text{ContactEvaluation}(\boldsymbol{u}^i_{n})$ \tcp{Same as Algorithm~\ref{Alg:dynamics}} 
  Calculate initial $\mathbf{R}^i_{\text{total}}(\boldsymbol{u}^i)$:
  $\leftarrow \boldsymbol{F}^i_{\text{int}} - \boldsymbol{F}^i_{\text{ext}}(\boldsymbol{u}^i) - \boldsymbol{F}^i_{\text{cont}}$ \tcp{Same as Algorithm~\ref{Alg:dynamics}}
  \While{$\|\mathbf{R}^i_{\mathrm{total}}\| > \epsilon_{\mathrm{inner}}$}{
    Solve linear system: $\mathbf{K}^i \Delta\boldsymbol{u}^i = -\mathbf{R}^i_{\mathrm{total}}$ \tcp{See Eq.~\eqref{eq:newton_iteration}}
    Update displacement: $\boldsymbol{u}^i \leftarrow \boldsymbol{u}^i + \Delta\boldsymbol{u}^i$ \tcp{See Eq.~\eqref{eq:update}}
    Update $\mathbf{R}^i_{\text{total}}(\boldsymbol{u}^i)$ \tcp{Recompute residual with updated $\boldsymbol{u}^i$}
  }
  $n \leftarrow n + 1$
}
\KwOutput{Displacement $\boldsymbol{u}^i$}
\end{algorithm}
Notably, a two-loop (an outer fixed-point loop with an inner Newton-Raphson loop) iteration is employed in this work, rather than a monolithic Newton scheme, primarily because the direct Newton method requires the contact Jacobian matrix $\mathbf{K}^i_{\text{cont}} = \frac{\partial \boldsymbol{F}^i_{\text{cont}}}{\partial \boldsymbol{u}^i}$ (second-order derivative of the contact energy function), which is not directly accessible from the current FMC-based contact force estimation. 
The fixed-point iteration scheme effectively handles this limitation by avoiding the explicit need for the Jacobian. 
To achieve higher-order derivatives, it is postulated that higher-order geometric primitives (e.g., surface patches) beyond fibers must be used, which incorporate not only location and extent, but also curvature information.
The development of such higher-order Monte Carlo methods is highly desired and should be considered as promising future work.

\section{Numerical examples}
\label{Sec:examples}
In this section, we demonstrate the capability of the proposed computational contact framework through a series of numerical examples of increasing complexity. The selected cases are carefully designed to verify both the accuracy and flexibility of the method. These cases range from classical small-deformation contact problems, e.g., Hertzian contact, which serves as a benchmark for the fundamental Signorini problem~\cite{laursen2003computational}, to more challenging finite-deformation contact scenarios involving severe geometric and material nonlinearities. We begin by comparing the method with the analytical solution for small-strain problems to demonstrate its accuracy in predicting basic contact mechanics responses. Subsequently, we assess the performance of the proposed method in large-deformation regimes using Neo-Hookean hyperelastic materials to demonstrate its effectiveness in modeling complex, nonlinear mechanical responses. Collectively, these examples provide a comprehensive assessment of the performance of our computational framework in addressing various contact problems characterized by nonlinear material behavior and complex geometric configurations.

\subsection{Small-strain contact problem}
\label{Sec:example:small}
In this section, we demonstrate the accuracy of the proposed contact formulation in the small-deformation regime against three canonical Signorini-type benchmark problems. These benchmarks include variations in geometry and loading conditions to comprehensively assess the performance of our method.

\subsubsection{Elastic semicircle–rigid surface contact}
\label{Sec:example:semicirlce}
\begin{figure}[H] \centering
    \subfigure[]
    {\includegraphics[width=0.47\textwidth]{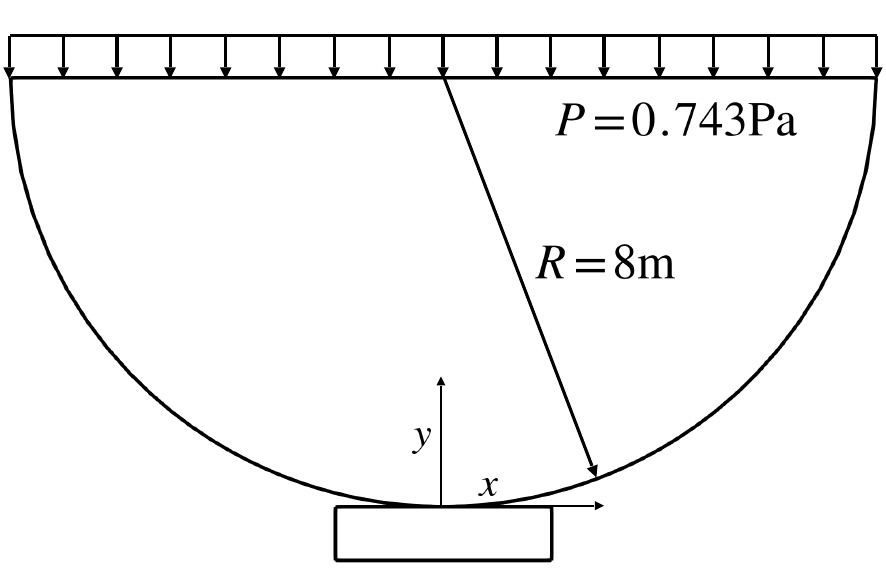}\label{fig:2.1.1}}
    \hfill
    \subfigure[]
    {\includegraphics[width=0.47\textwidth]{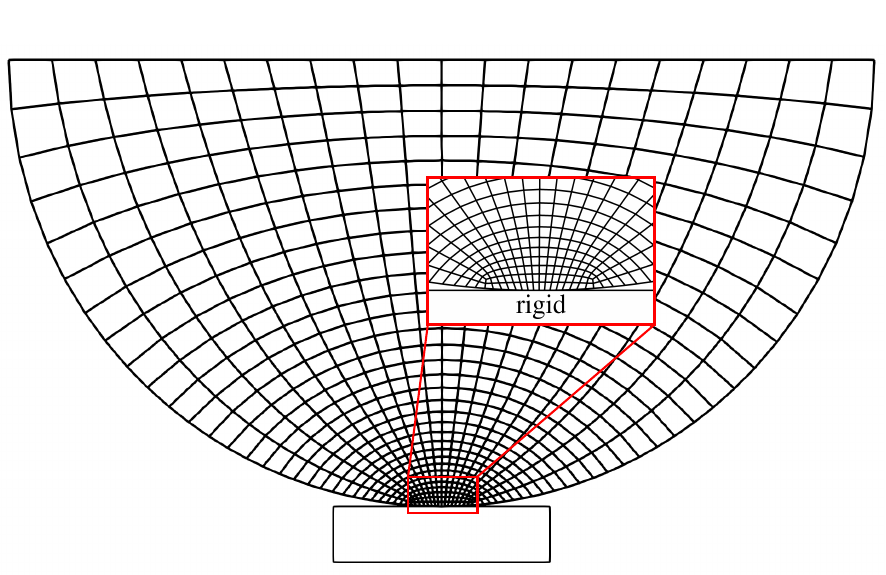}\label{fig:2.1.2}} 
    \caption{The model setup: (a) geometry and boundary conditions. (b) finite element mesh.} \label{fig:2.1}
\end{figure}

The Hertzian contact problem is a classical benchmark in contact mechanics due to the availability of closed-form solutions for non-conformal smooth surfaces in contact. We consider the contact between a deformable semicircular cylinder of radius $R=8~\mathrm{m}$ and a rigid plane under plane-strain conditions, as illustrated in Fig.~\ref{fig:2.1}\subref{fig:2.1.1}. The upper surface of the semicircle is subjected to a uniformly distributed pressure $p$. The cylinder is modeled as a linear elastic body undergoing infinitesimal deformations, with its constitutive relation given by  $\boldsymbol{\sigma} = \lambda \, \mathrm{tr}(\boldsymbol{\varepsilon}) \, \boldsymbol{I} + 2\mu \, \boldsymbol{\varepsilon}$, where $\lambda$ and $\mu$ are the Lamé constants, corresponding to Young’s modulus $E=200~\mathrm{Pa}$ and Poisson’s ratio $\nu = 0.3$. According to Hertzian contact theory, the normal contact pressure distribution $p(x)$ and the half-width of the contact area $a$ are given by~\cite{liang2024mortar}:

\begin{subequations}\label{eq_semicircle}
\begin{align}
p_0 &= \frac{4Rp}{\pi b^2} \sqrt{a^2 - x^2}, \\
a &= 2 \sqrt{\frac{2R^2 p (1 - \nu^2)}{E \pi}},
\end{align}
\end{subequations}
where $x$ is the coordinate along the contact surface, with the origin located at the center of the contact region. 

\begin{figure}[H] \centering
    {\includegraphics[width=0.7\textwidth]{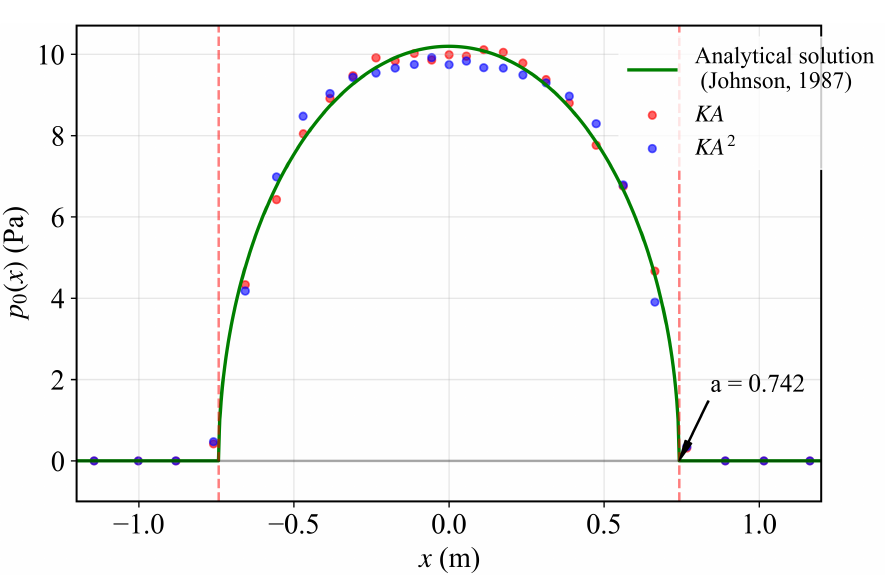}}
    \caption{Comparison of contact pressure distribution between numerical simulation and analytical solution~\cite{johnson1987contact}.} \label{fig:2.2}
\end{figure}

\begin{figure}[H] \centering
    \subfigure[]
    {\includegraphics[width=0.49\textwidth]{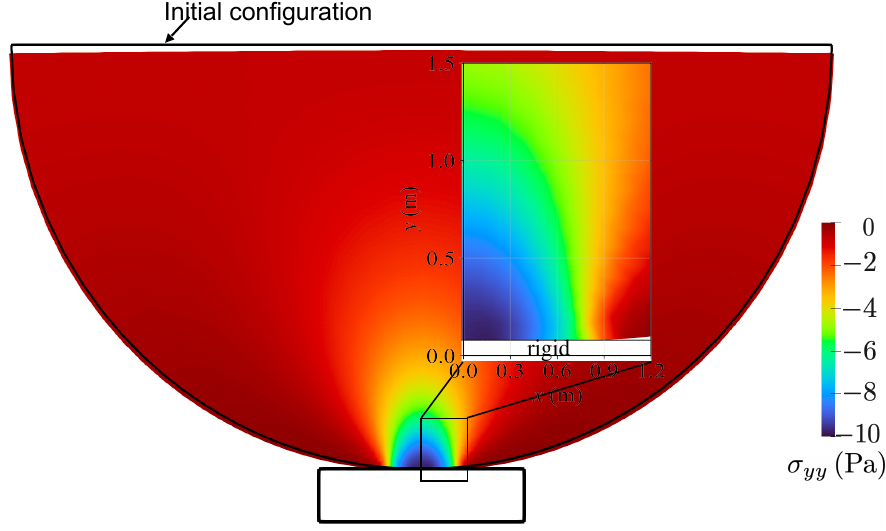}\label{fig:2.3.1}}
    \hfill
    \subfigure[]
    {\includegraphics[width=0.487\textwidth]{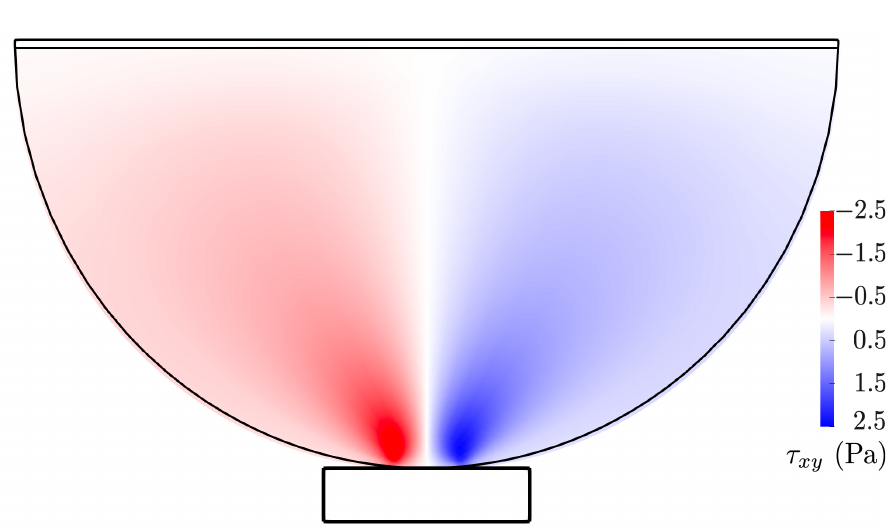}\label{fig:2.3.2}} 
    \caption{Distributions of (a) vertical stress $\sigma_{yy}$ and (b) shear stress $\tau_{xy}$, represented by colormaps.} \label{fig:2.3}
\end{figure}

To accurately capture the localized deformation near the contact zone, the mesh is refined in the vicinity of the contact interface, as illustrated in Fig.~\ref{fig:2.1}\subref{fig:2.1.2}. A comparison between the numerically computed and analytical contact pressure distributions is presented in Fig.~\ref{fig:2.2}. As introduced in Eq.~\eqref{eq_energy_function}, the proposed contact formulation is derived from an energy-based function \( w(A) \) that governs the interaction between contacting bodies through the overlap measure \( A \). To assess the generality of this energy model in our framework, we consider both a linear potential \( w = K A \) and a quadratic potential \( w = K A^2 \). Both forms accurately reproduce the Hertzian pressure distribution. This result demonstrates that the framework is general, with its accuracy independent of the specific form of the energy function. Despite minor deviations near the edges of the contact region, attributable to mesh resolution and discretization effects, the numerical results exhibit good agreement with the Hertzian analytical solution. The corresponding distributions of vertical stress \( \sigma_{yy} \) and shear stress \( \tau_{xy} \) are shown in Fig.~\ref{fig:2.3}, respectively. In particular, the shear stress field in Fig.~\ref{fig:2.3}\subref{fig:2.3.2} remains continuous and symmetric across the contact area, consistent with theoretical expectations of localized stress concentration at the interface. This benchmark problem quantitatively confirms the accuracy of the proposed contact formulation in reproducing both the pressure distribution and the internal stress response under classical small-strain Hertzian contact conditions.

\subsubsection{Indentation by a wedge}
\label{Sec:example:wedge}
\begin{figure}[H] \centering
    \subfigure[]
    {\includegraphics[width=0.48\textwidth]{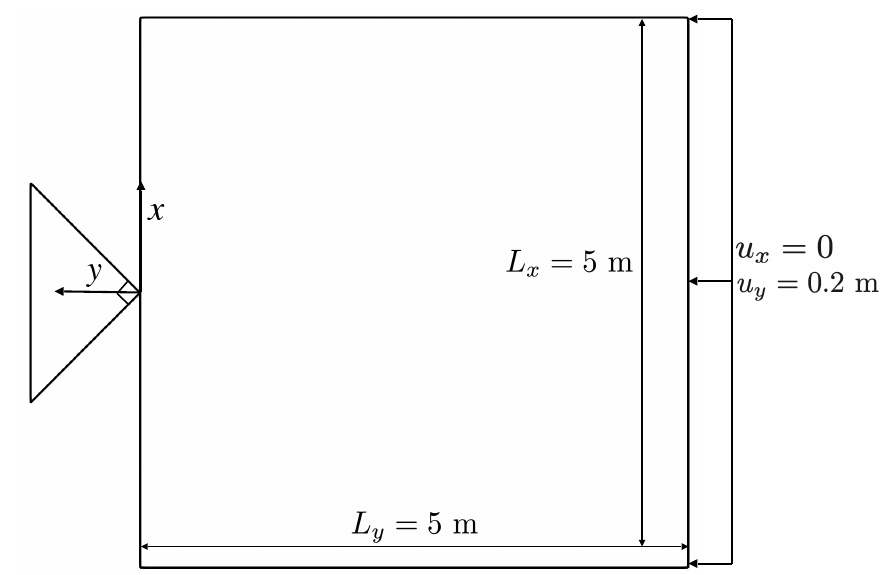}\label{fig:2.4.1}}
    \hfill
    \subfigure[]
    {\includegraphics[width=0.48\textwidth]{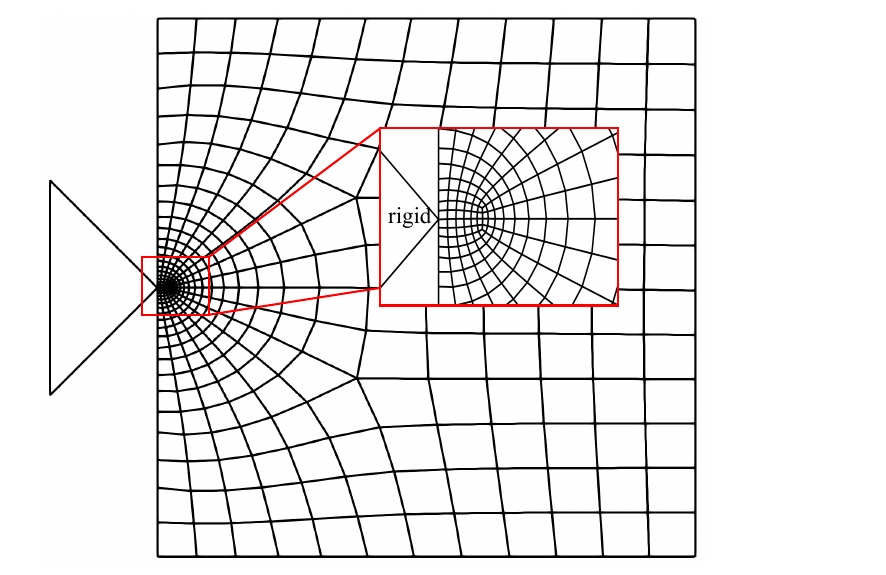}\label{fig:2.4.2}} 
    \caption{The model setup: (a) geometry and boundary conditions, and (b) discretized mesh.} \label{fig:2.4}
\end{figure}

As described in Section~\ref{sec:intro}, contact analyses involving sharp corners suffer from geometric singularities with ill-defined or discontinuous surface normals, posing challenges for traditional algorithms. To demonstrate the effectiveness of the proposed scheme, we consider a 2D rigid wedge indentation scenario. The simulation setup is shown in Fig.~\ref{fig:2.4}. The rigid wedge is held fixed while the elastic block is moved leftwards to create the indentation. This motion is achieved by prescribing a horizontal displacement along the right boundary of the block (y-axis direction) up to a maximum of $4\%$ of the block's length, while keeping the vertical displacement fixed at this boundary. The top and bottom boundaries of the block remain traction-free. The elastic block is modeled as a homogeneous, isotropic linear elastic material ($E = 500~\mathrm{Pa}$, $\nu = 0.3$) under small-deformation assumptions. Given the small-deformation regime, a linear elastic model is employed.

The analytical solution~\cite{popov2025handbook} for the contact pressure distribution $p$ and the total contact force $P$ is given by:
\begin{subequations}\label{eq_wedge1}
\begin{align}
p(x;a) &= 
\begin{cases}
\displaystyle \frac{E^* \tan(\beta)}{\pi} \, \mathrm{artanh} \left( \sqrt{1 - \frac{x^2}{a^2}} \right), & 0 < |x| \leq a, \\[8pt]
0, & |x| > a,
\end{cases} \label{eq_wedge1a} \\
P(a) &= E^* \tan(\beta)a, \label{eq_wedge1b}
\end{align}
\end{subequations}
where $a$ is the half-width of the contact area, $E^* = E / (1 - \nu^2)$ is the equivalent elastic modulus under plane strain conditions, and $\beta$ is the slope angle of the wedge, set to $45^\circ$ in this case. The coordinate $x$ is defined along the contact interface, with the origin located at the center of the contact region. 

The horizontal position of the deformed surface, denoted as $h(x; a)$, is defined as the height measured from the tip of the rigid wedge along the horizontal direction~\cite{popov2025handbook}. By this definition, $h(x; a)$ represents the gap between the wedge tip and the deformed surface: it is zero at the center of contact ($h(0; 0) = 0$) and increases linearly to a maximum $a \tan(\beta)$ at the boundaries ($x = \pm a$). The complete expression of $h(x; a)$ is given by

\begin{align}\label{eq_wedge2}
h(x; a) = 
\begin{cases}
\tan(\beta) \cdot |x|, & 0 < |x| \leq a,\\[8pt]
\displaystyle \frac{2\tan(\beta)}{\pi} \left[ a \, \mathrm{arcosh} \left( \frac{|x|}{a} \right) + |x| \, \mathrm{arcsin} \left( \frac{a}{|x|} \right) \right], & |x| \geq a.
\end{cases}
\end{align}

\begin{figure}[H] \centering
    \subfigure[]
    {\includegraphics[width=0.49\textwidth]{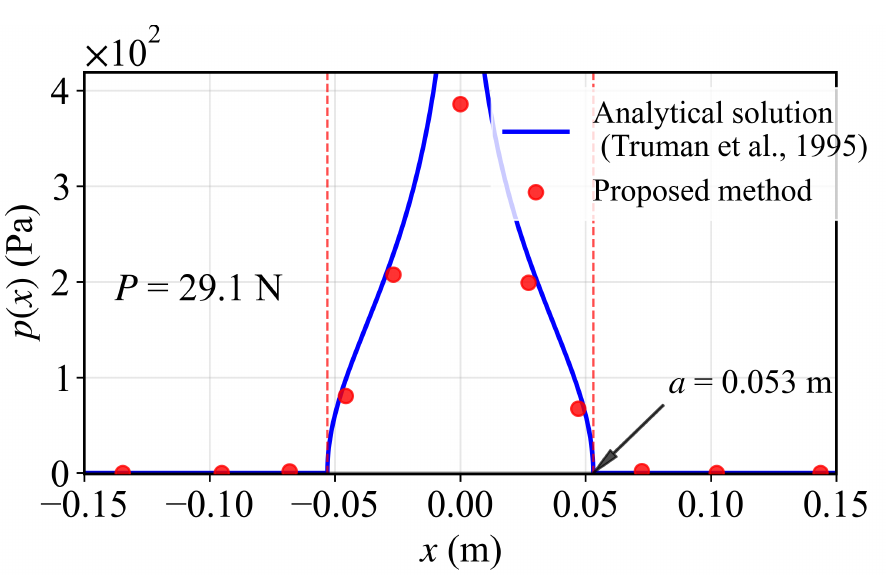}\label{fig:2.5.1}}
    \hfill
    \subfigure[]
    {\includegraphics[width=0.49\textwidth]{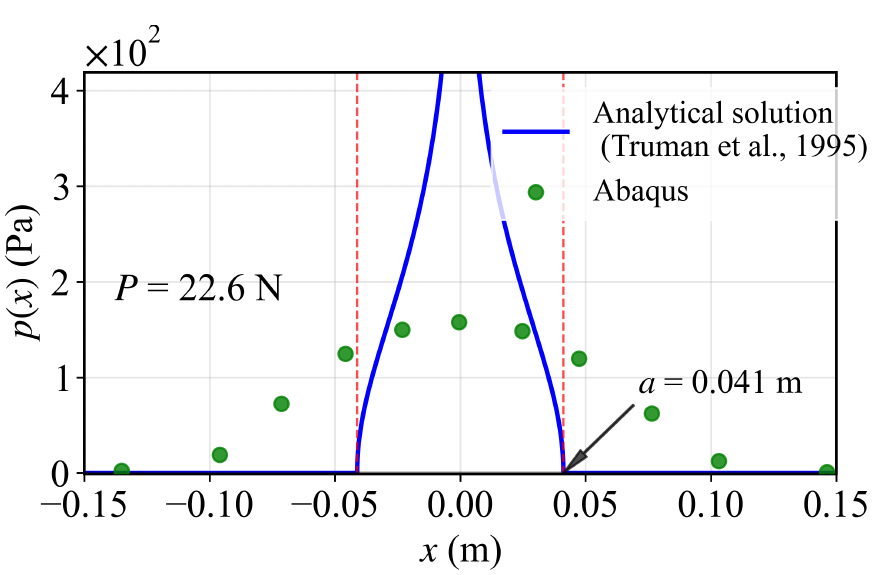}\label{fig:2.5.2}} 
    \caption{Comparison of contact pressure distribution between numerical simulation and analytical solution~\cite{truman1995contact}: (a) the present scheme, and (b) Abaqus.} \label{fig:2.5}
\end{figure}

\begin{figure}[H] \centering
    \subfigure[]
    {\includegraphics[width=0.85\textwidth]{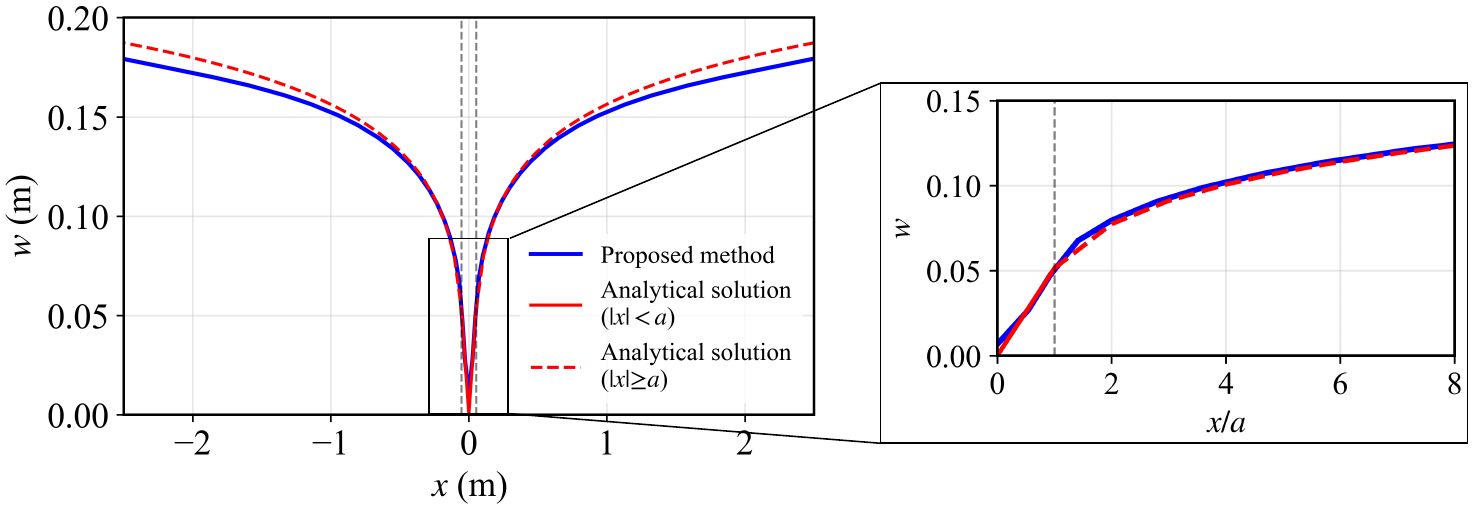}\label{fig:2.6.1}}
    \\
    \subfigure[]
    {\includegraphics[width=0.85\textwidth]{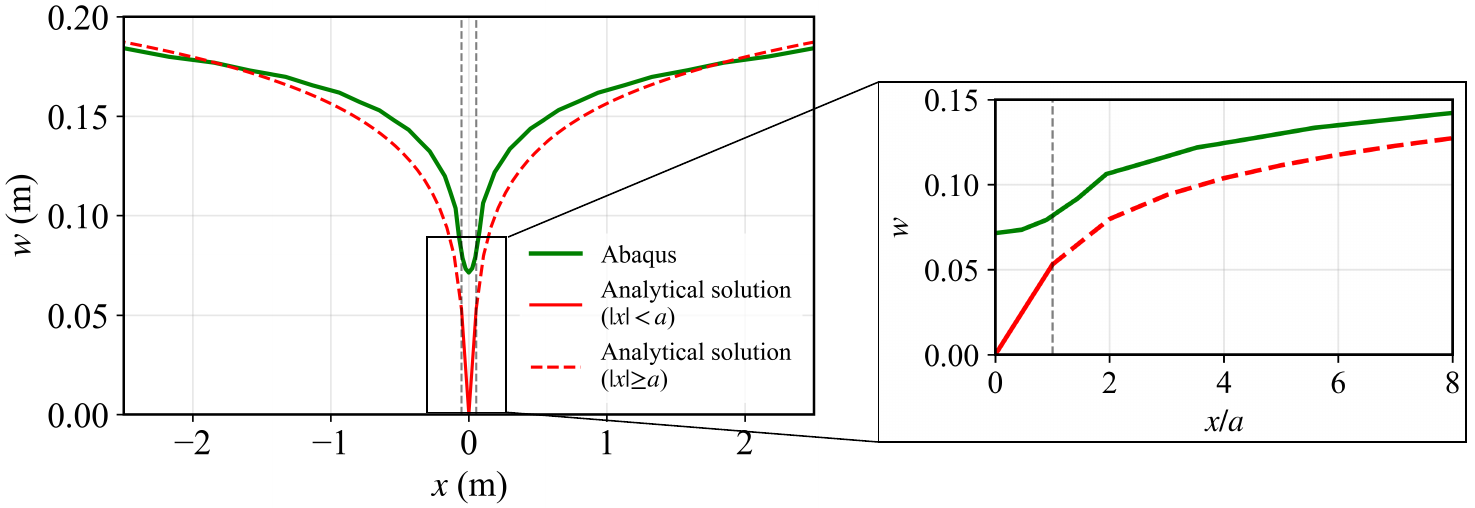}\label{fig:2.6.2}} 
    \caption{Comparison of displacement distribution between numerical simulation and analytical solution~\cite{truman1995contact}: (a) the present scheme, and (b) Abaqus.} \label{fig:2.6}
\end{figure}

\begin{figure}[H] \centering
    {\includegraphics[width=0.8\textwidth]{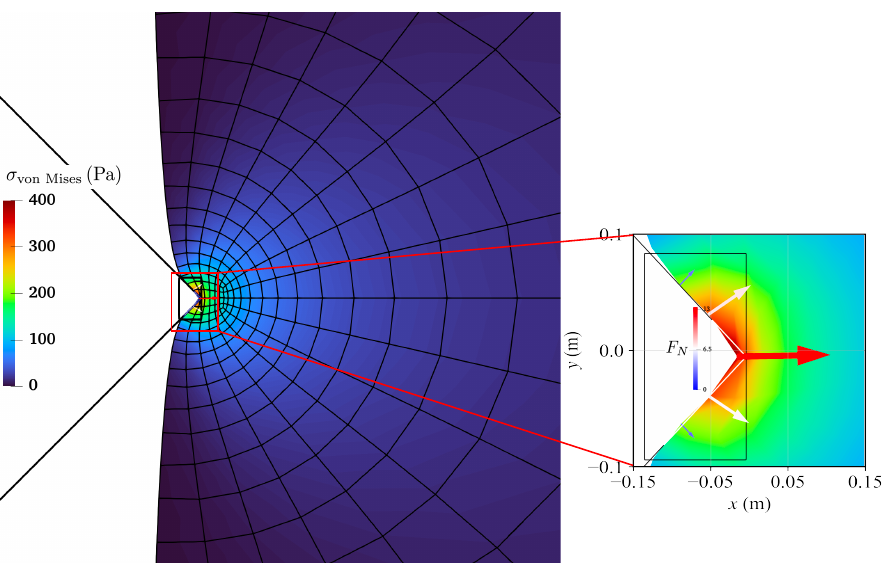}}
    \caption{Distribution of von mises stress $\sigma_{\mathrm{von\ Mises}}$, represented by colormap.} \label{fig:2.7}
\end{figure}

For contact between a rigid wedge indenter and an elastic block, numerical simulations were conducted using both the proposed framework and Abaqus. The contact half-width $a$ is derived from the contact force $P$ via Eq.~\eqref{eq_wedge1b}, where $P$ is obtained separately from the Abaqus simulations and our proposed method. The numerical results from our method and Abaqus are compared against the analytically derived pressure distribution in Figs.~\ref{fig:2.5}\subref{fig:2.5.1} and \subref{fig:2.5.2}, respectively. As illustrated in Fig.~\ref{fig:2.5}\subref{fig:2.5.1}, the pressure distribution obtained from the proposed framework closely matches the analytical solution. In contrast, Fig.~\ref{fig:2.5}\subref{fig:2.5.2} shows that the Abaqus results deviate noticeably from the analytical solution, particularly in terms of the width $a$ of the contact region. Fig.~\ref{fig:2.6} presents the deformation profiles of the contact surface obtained from our method and Abaqus, each compared against the corresponding analytical solution. The proposed method produces results that align closely with the analytical solution, as seen in Fig.~\ref{fig:2.6}\subref{fig:2.6.1}. Only minor numerical errors are observed in the far-field region, likely due to the finite mesh resolution. In contrast, the Abaqus solution exhibits significant over-penetration, as shown in Fig.~\ref{fig:2.6}\subref{fig:2.6.2}, likely arising from an underestimation of the contact force $P$ in the simulation. The von Mises stress distribution in the elastic block is presented in Fig.~\ref{fig:2.7}, with an inset highlighting the magnitude and direction of the contact force via arrow-scaled contours. Overall, this numerical example demonstrates the better accuracy of the proposed contact scheme over conventional methods (e.g., the default penalty method in Abaqus) in simulating challenging contact scenarios involving sharp geometric features, which often induce ill-defined situations in contact detection and normal direction for traditional node-to-surface or surface-to-surface contact algorithms, leading to numerical inaccuracy.

\subsubsection{Indentation by a cone}
\label{Sec:example:cone}
\begin{figure}[H] \centering
    {\includegraphics[width=0.8\textwidth]{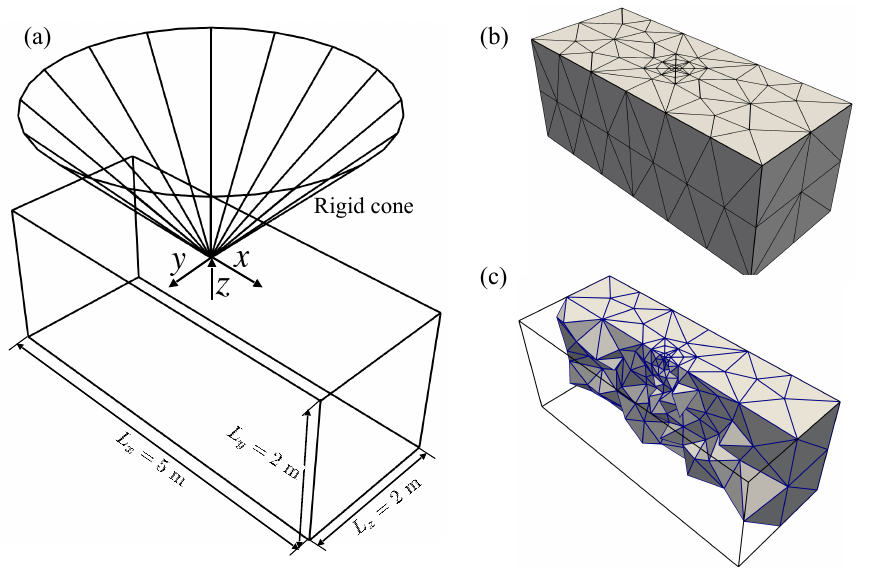}}
    \caption{The model setup: (a) geometry, (b) mesh, and (c) clipped view of tetrahedral mesh.} \label{fig:2.8}
\end{figure}

To examine the applicability of the proposed contact scheme in three-dimensional scenarios, we consider an indentation problem involving a rigid conical indenter and an elastic square block, as illustrated in Fig.~\ref{fig:2.8}(a).  The rigid conical indenter introduces geometric nonsmoothness, posing challenges for traditional contact algorithms. This numerical example serves to examine the capability of the proposed framework in handling three-dimensional contact with nonsmooth geometry, including accurate prediction of pressure distribution and indentation profile, under small-strain linear elastic conditions. The 3D domain is discretized with tetrahedral elements, with illustrations of the complete and clipped meshes shown in Fig.~\ref{fig:2.8}(b) and (c), respectively. The bottom surface of the block is subjected to a prescribed vertical displacement $u_z = 0.2\,\mathrm{m}$, while lateral (horizontal) displacements are constrained to prevent rigid-body motion. The block material is linear elastic with parameters $E = 500\,\mathrm{Pa}$ and $\nu = 0.35$, and small deformations are assumed. 

According to~\cite{popov2019handbook}, the normal contact pressure $p$ and the indentation profile $h$ are given by:
\begin{subequations}\label{eq_cone}
\begin{align}
p(r; a) &=
\begin{cases}
-\dfrac{E^* d}{\pi a} \mathrm{arcosh}\left(\dfrac{a}{r}\right), & r \leq a, \\[2mm]
0, & r > a,
\end{cases} \\
h(r; a) &=
\begin{cases}
\tan(\beta) \cdot r, & r \leq a, \\[2mm]
a \tan(\beta) \left[ \arcsin\left(\dfrac{a}{r}\right) + \dfrac{\sqrt{r^2 - a^2} - r}{a} \right], & r > a,
\end{cases}
\end{align}
\end{subequations}
where $a$ is the contact radius (half-width of the contact region). The corresponding indentation depth $d$ and normal contact force $F_N$ are given by:

\begin{equation}
d = \frac{\pi}{2} a \tan \beta, \quad F_N = a E^* d,
\end{equation}
where $\beta$ is the slope angle of the cone, and $r$ is the radial coordinate measured from the center of the contact region.

\begin{figure}[H] \centering
    \subfigure[]
    {\includegraphics[width=0.5\textwidth]{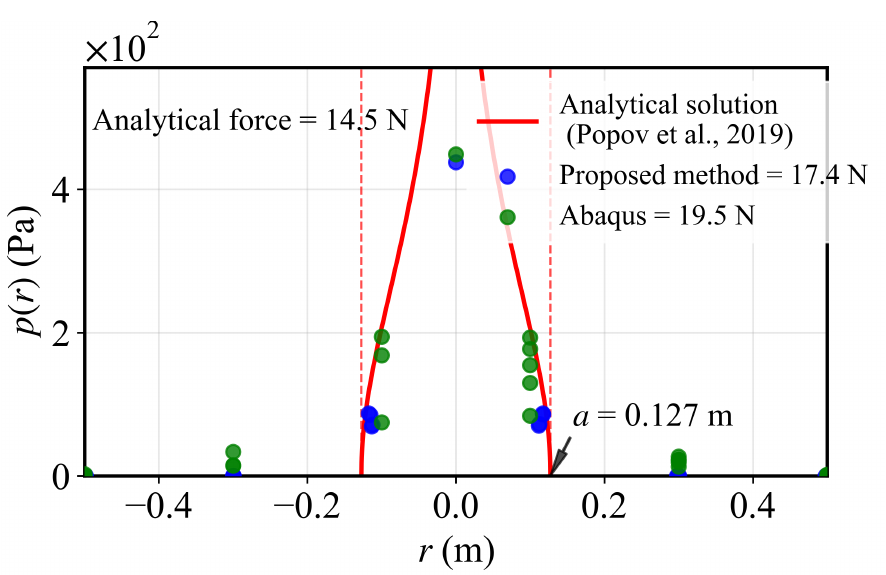}\label{fig:2.9.1}}
    \hfill
    \subfigure[]
    {\includegraphics[width=0.48\textwidth]{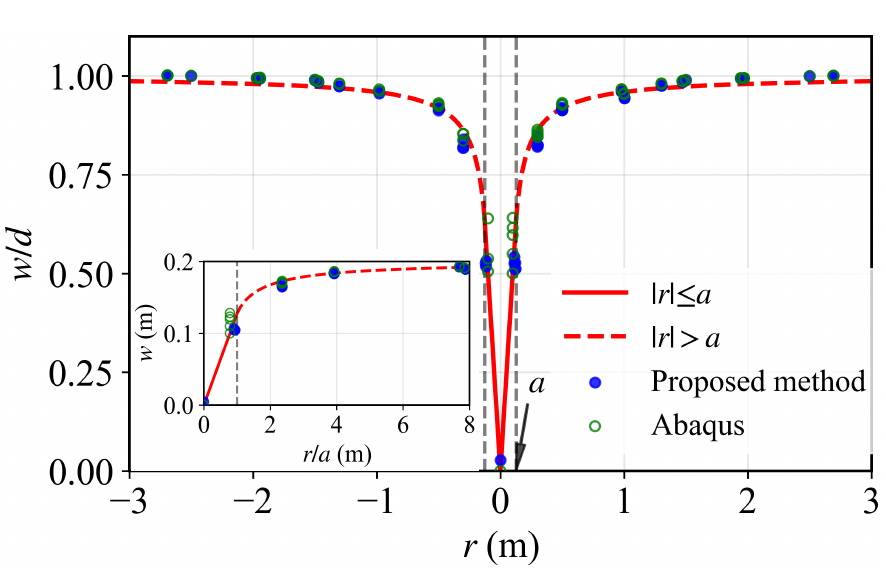}\label{fig:2.9.2}} 
    \caption{Comparison of contact pressure and displacement distributions among our proposed method, Abaqus simulations, and analytical solutions~\cite{popov2019handbook}: (a) pressure distribution; (b) displacement distribution.} \label{fig:2.9}
\end{figure}

\begin{figure}[H] \centering
    {\includegraphics[width=1\textwidth]{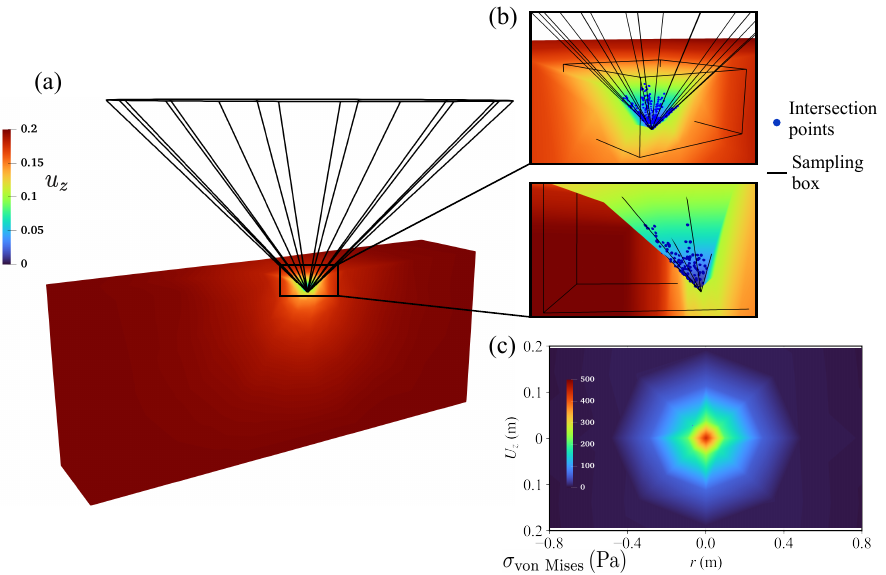}}
    \caption{Visualization of (a) displacement field $u_z$, (b) deformed configuration near the contact interface with intersection markers indicating the points where fibers cross the contact boundary, and (c) distribution of von Mises stress field $\sigma_{\mathrm{von\ Mises}}$.} \label{fig:2.10}
\end{figure}

Fig.~\ref{fig:2.9}\subref{fig:2.9.1} compares the contact pressure distributions along the radial direction \(r\), measured from the center of the contact region, as predicted by the proposed computational contact framework, Abaqus, and the analytical solution. Under the given mesh discretization, the contact force predicted by the present scheme is $17.4\,\mathrm{N}$, which is closer to the reference solution of $14.5\,\mathrm{N}$ than the Abaqus result of $19.5\,\mathrm{N}$. This comparison demonstrates the better accuracy of the proposed framework in predicting contact forces. Moreover, a comparison of the results in Fig.~\ref{fig:2.9}\subref{fig:2.9.1} demonstrates that the contact pressure distribution predicted by the proposed method exhibits improved symmetry and uniformity compared to the results obtained by
conventional Abaqus simulation, particularly at radial positions corresponding to $r$. 
The deformation profiles obtained from both numerical schemes show good agreement with the analytical solution, as depicted in Fig.~\ref{fig:2.9}\subref{fig:2.9.2}. Fig.~\ref{fig:2.10} presents additional visualizations: (a) Distribution of the \(z\)-direction displacement of the entire elastic body under indentation, highlighting the global deformation pattern represented by color map; (b) a detailed view of the sampling region in the vicinity of the contact interface, illustrating the points where the fibers intersect the contact boundary; (c) a zoomed-in side view of the sampling region, revealing the deformed profile of the elastic body and the points at which the fibers intersect the contact boundary; and (d) the von Mises stress distribution in the elastic body, plotted with respect to radial distance and vertical displacement $u_z$. In particular, the blue dots in the local sampling regions (Fig.~\ref{fig:2.10}(b) and (c)) indicate the intersection points between the fibers and the contact surface. The distribution of von Mises stress demonstrates that the proposed method captures a smooth and symmetric stress distribution throughout the contact area, confirming the capability of the framework to handle three-dimensional contact problems with nonsmooth geometry accurately.

\subsection{Finite-strain hyperelastic contact problem}
\label{Sec:example:large}
This section investigates contact scenarios involving large elastic deformations, further demonstrating the robustness and generality of the proposed computational framework. In these examples, the material behavior is modeled using a classical Neo-Hookean constitutive law derived from the strain energy function:

\begin{subequations}\label{eq_neoH}
\begin{align}
W(\boldsymbol{F}) &= \frac{G}{2} \left( J^{-2/3} I_1 - 3 \right) + \frac{K}{2} (J - 1)^2, \\
J &= \det(\boldsymbol{F}), \\
I_1 &= \mathrm{tr}(\boldsymbol{C}),
\end{align}
\end{subequations}
where $\boldsymbol{F}$ is the deformation gradient. The first invariant $I_1$ is derived from the right Cauchy–Green tensor $\boldsymbol{C}$, which is defined as $\boldsymbol{C} = \boldsymbol{F}^\top \boldsymbol{F}$. The material constants $G$ and $K$ represent the shear and bulk moduli, respectively. These moduli can be expressed in terms of Young's modulus $E$ and Poisson's ratio $\nu$ as $G = E / [2(1 + \nu)]$ and $K = E / [3(1 - 2\nu)]$.

The numerical implementation follows the same framework introduced in Sec.~\ref{sec:computational framework} and is extended here to handle finite-deformation contact problems involving hyperelastic materials. These examples allow for a systematic evaluation of the proposed framework under nonlinear geometric and material conditions, further demonstrating its applicability beyond the small-strain linear elasticity regime.

\subsubsection{Ironing test}
\label{Sec:example:iron}
In this section, we consider two related benchmark scenarios to assess the proposed contact scheme under finite-strain conditions. 
The first case involves an elastic flat punch pressed vertically onto an elastic block, serving to verify the accuracy of our framework by comparing its results with established references under large-deformation compression. Following this assessment, the second case extends the setup by including horizontal sliding along the top surface of the block in addition to vertical compression, thereby assessing the performance of the proposed method under combined large-deformation and sliding contact.

\begin{figure}[H] \centering
    {\includegraphics[width=0.7\textwidth]{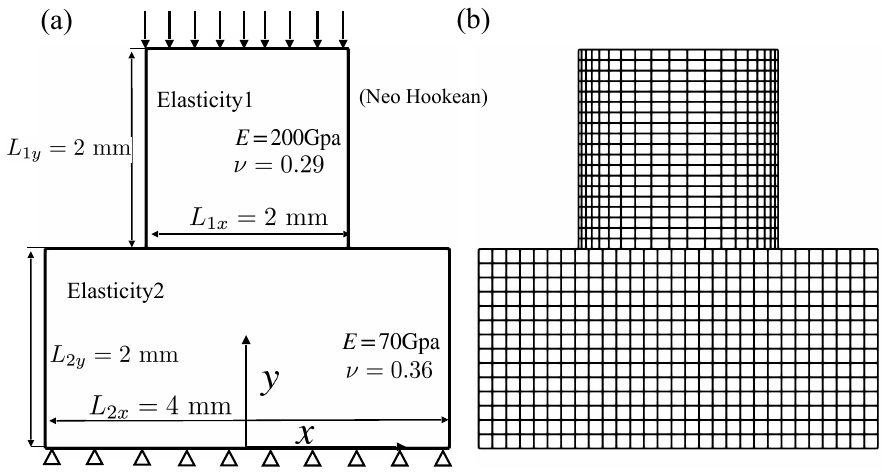}}
    \caption{A schematic model of a flat punch on a block: (a) geometry and boundary conditions, and (b) discretized mesh.} \label{fig:2.11}
\end{figure}

\begin{figure}[H] \centering
    {\includegraphics[width=0.75\textwidth]{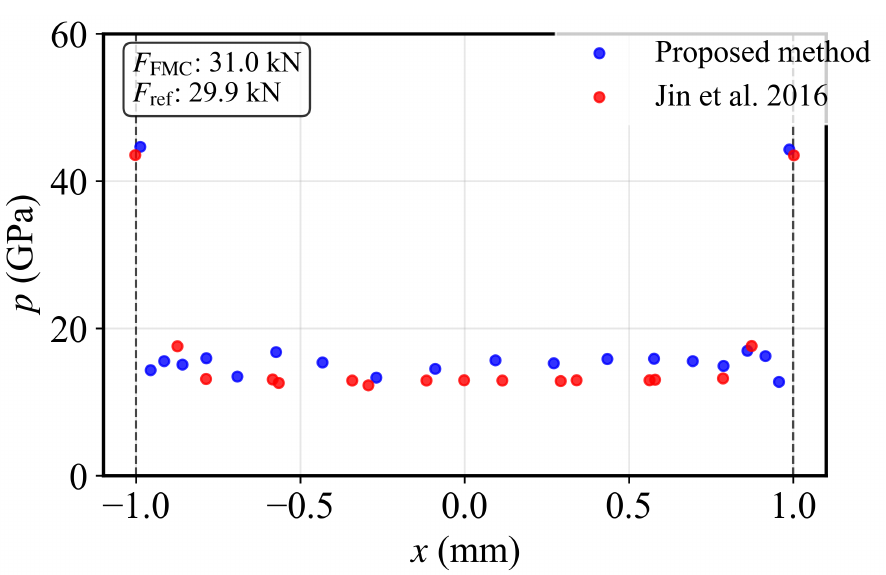}}
    \caption{ Comparison of contact pressures obtained from the present scheme and reference~\cite{jin2016node}.} \label{fig:2.12}
\end{figure}

For the first scenario, an elastic flat punch is pressed vertically onto an elastic block by imposing a displacement of $0.4\,\mathrm{mm}$ on its top surface, while the block is fixed in the $x$ and $y$ directions on the bottom surface. The bottom surface of the punch contacts with the block, which is free to deform, and the overall geometry and mesh discretization are illustrated in Fig.~\ref{fig:2.11}. The material properties of the punch and the block  are assigned as follows: the punch has $E = 200\,\mathrm{GPa}$, $\nu = 0.29$, and the block has $E = 70\,\mathrm{GPa}$, $\nu = 0.36$. Fig.~\ref{fig:2.12} presents the contact pressure distribution at the punch–block interface computed by the proposed scheme, compared with reference results from~\cite{jin2016node}. The numerical solution reproduces the main features of the pressure profile reported in the literature, including the peak contact pressure and its spatial variation, demonstrating the accuracy of the proposed framework in capturing nonlinear deformations. 

\begin{figure}[H] \centering
    {\includegraphics[width=0.7\textwidth]{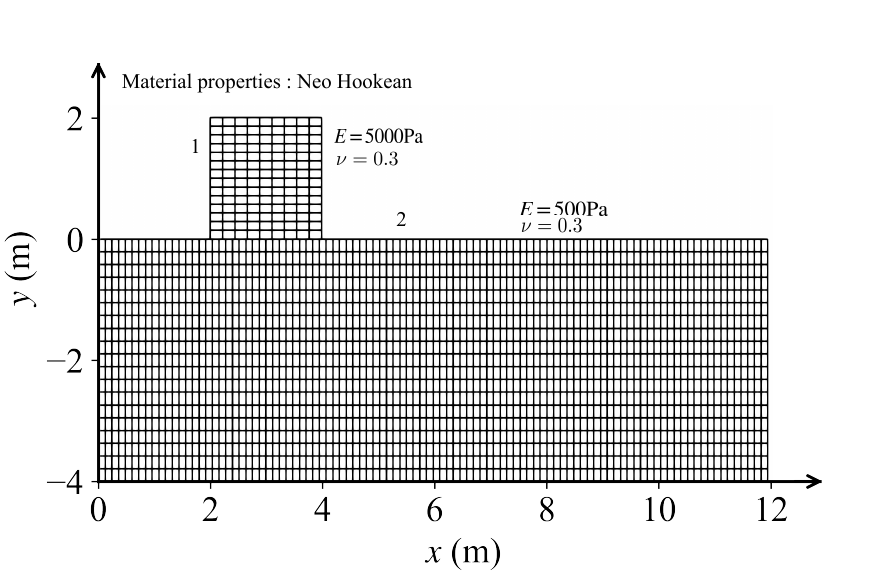}}
    \caption{A schematic model of the ironing of an elastic foundation.} \label{fig:2.13}
\end{figure}

\begin{figure}[H] \centering
    {\includegraphics[width=1\textwidth]{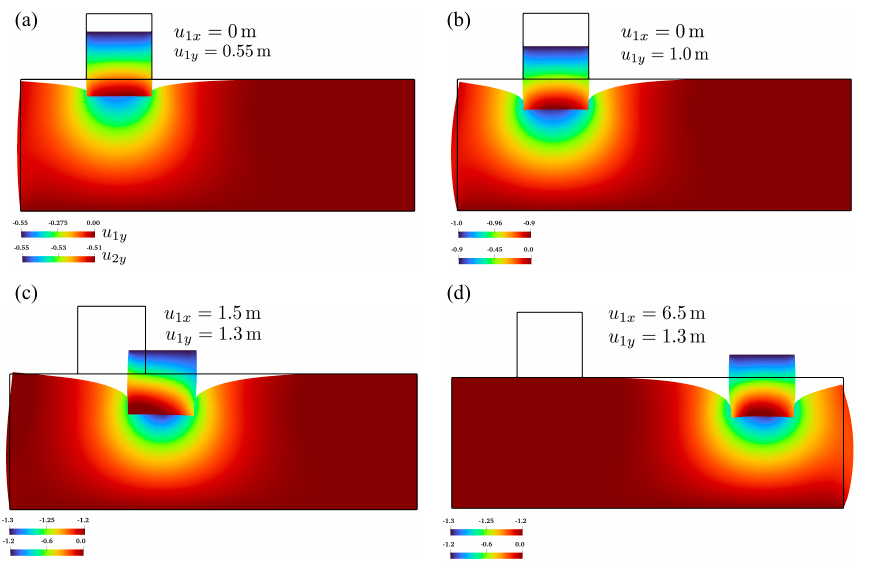}}
    \caption{Ironing of an elastic foundation: Reference configuration and current configurations with displacement distribution during the loading process: (a) step 4, (b) step 7, (c) step 11, and (d) step 21.} \label{fig:2.14}
\end{figure}

\begin{figure}[H] \centering
    {\includegraphics[width=0.65\textwidth]{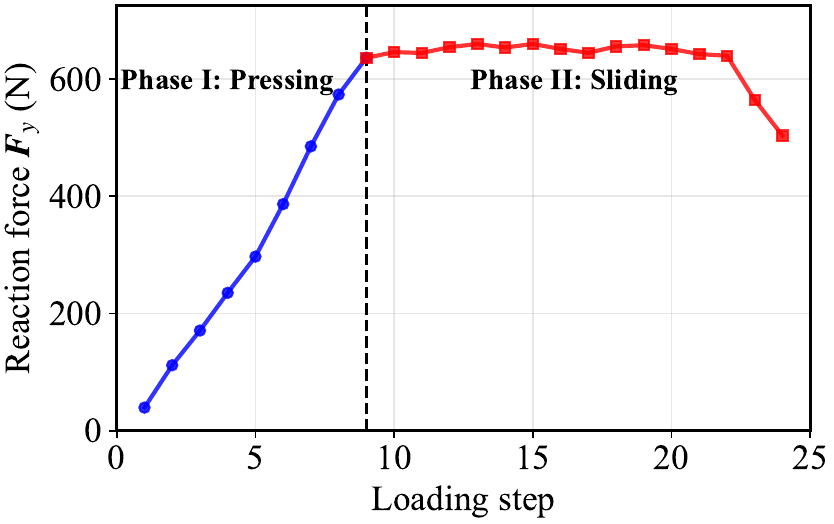}}
    \caption{Ironing problem: computed total vertical reaction force exerted at the indenter.} \label{fig:2.15}
\end{figure}

To further assess the performance of the proposed scheme under combined large-deformation and sliding contact, we introduce a dynamic sliding problem following the static indentation benchmark. A stiffer elastic indenter ($E = 5000\,\mathrm{Pa}$, $\nu = 0.3$) is initially pressed onto a softer elastic foundation ($E = 500\,\mathrm{Pa}$, $\nu = 0.3$), after which it slides horizontally along the top surface of the block. The geometry and dimensions of the problem are illustrated in Fig.~\ref{fig:2.13}. The vertical indentation is applied through a displacement of $v_y = 1.3\,\mathrm{m}$ on the top surface of the intender, incremented quasi-statically in eight equal steps of $\delta v_y = 0.225\,\mathrm{m}$ in the numerical simulation. After reaching the prescribed maximum vertical displacement, the indenter is driven horizontally for a total distance of $u_x = 7.5\,\mathrm{m}$, applied in fifteen equal increments of $\Delta u_x = 0.5\,\mathrm{m}$, with the vertical compression maintained. The reference configuration and selected deformed configurations are shown in Fig.~\ref{fig:2.14}, where subfigure (a) corresponds to step 4 and (b) to step 7 of the vertical indentation phase, and subfigure (c) corresponds to step 11 and (d) to step 21 of the subsequent sliding phase. The figure clearly distinguishes the global deformed configuration of the block. The evolution of the vertical and horizontal reaction forces is presented in Fig.~\ref{fig:2.15}. Both forces evolve smoothly without visible numerical oscillations or spurious penetration, demonstrating the stability of the proposed computational framework in capturing large-deformation sliding contact.

\subsubsection{Collision of two elastic bodies}
\label{Sec:example:star}
\begin{figure}[H] \centering
    {\includegraphics[width=0.8\textwidth]{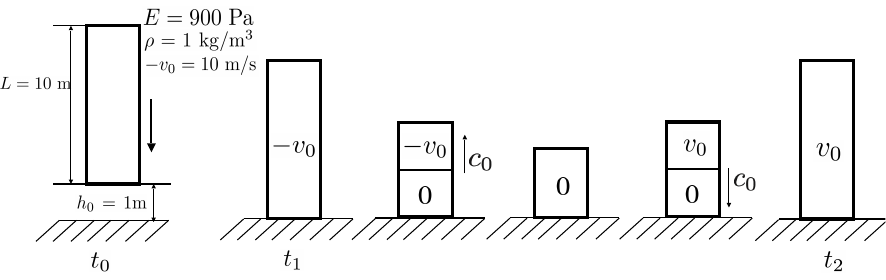}}
    \caption{One-dimensional benchmarks: impact of an elastic bar.} \label{fig:2.16}
\end{figure}

 To assess the performance of the proposed method in a dynamic contact setting, we begin with a simplified one-dimensional benchmark problem. This benchmark serves as a first test case to assess the stability and accuracy of the algorithm before we extend it to two-dimensional collisions with more complex geometries. The scenario involves an elastic bar impacting a rigid obstacle, as illustrated in Fig.~\ref{fig:2.16}. The bar has length $L = 10\,\mathrm{m}$, Young's modulus $E = 900\,\mathrm{Pa}$, density $\rho = 1\,\mathrm{kg/m^3}$, and gravity is neglected. The bar is initially undeformed and dropped from a height of $h_0 = 1\,\mathrm{m}$ with an initial velocity of $v_0 = 10\,\mathrm{m/s}$ toward the rigid obstacle located at $x = 0$. The top end of the bar ($x = L$) is subject to a zero-gradient (Neumann) boundary condition, representing a free end that allows stress waves to reflect without artificial constraint.  

The displacement $u(x,t)$ of the bar and the contact pressure $p(t)$ at $x=0$ satisfy the one-dimensional elastodynamic equations with unilateral contact constraints:

\begin{subequations}\label{eq_1D}
\begin{align}
&\rho \ddot{u} - E \frac{\partial^2 u}{\partial x^2} = p, \quad \text{in } \Omega \times (0,T), \\
&u(0,t) \geq 0, \; p(t) \geq 0, \; p(t) u(0,t) = 0 \quad \text{on } (0,T), \\
&\frac{\partial u}{\partial x}(L,t) = 0 \quad \text{on } (0,T), \\
&u(x,0) = h_0, \quad \dot{u}(x,0) = -v_0.
\end{align}
\end{subequations}
In this one-dimensional setting, the motion reflects the propagation of elastic waves along the bar, generating periodic stress accumulation and relaxation at the contact interface during impact. For the numerical simulation, the bar is discretized with $\Delta x = 0.1\,\mathrm{m}$, and the time step is chosen according to the CFL condition (CFL=0.5):
\begin{equation}
\mathrm{CFL} = \frac{c_0 \, \Delta t}{\Delta x}, \quad c_0 = \sqrt{\frac{E}{\rho}},
\end{equation}
where $c_0$ is the elastic wave speed. This choice ensures stability for the explicit time integration. During the simulation, the bar exhibits periodic rigid-body behavior: it impacts the obstacle, remains in contact for one time unit per cycle while stress waves propagate along its length, and subsequently rebounds when the reflected waves return to the contact zone. The propagation of stress waves within the bar is schematically illustrated by the arrows in Fig.~\ref{fig:2.16}. 

\begin{figure}[H] \centering
    {\includegraphics[width=0.9\textwidth]{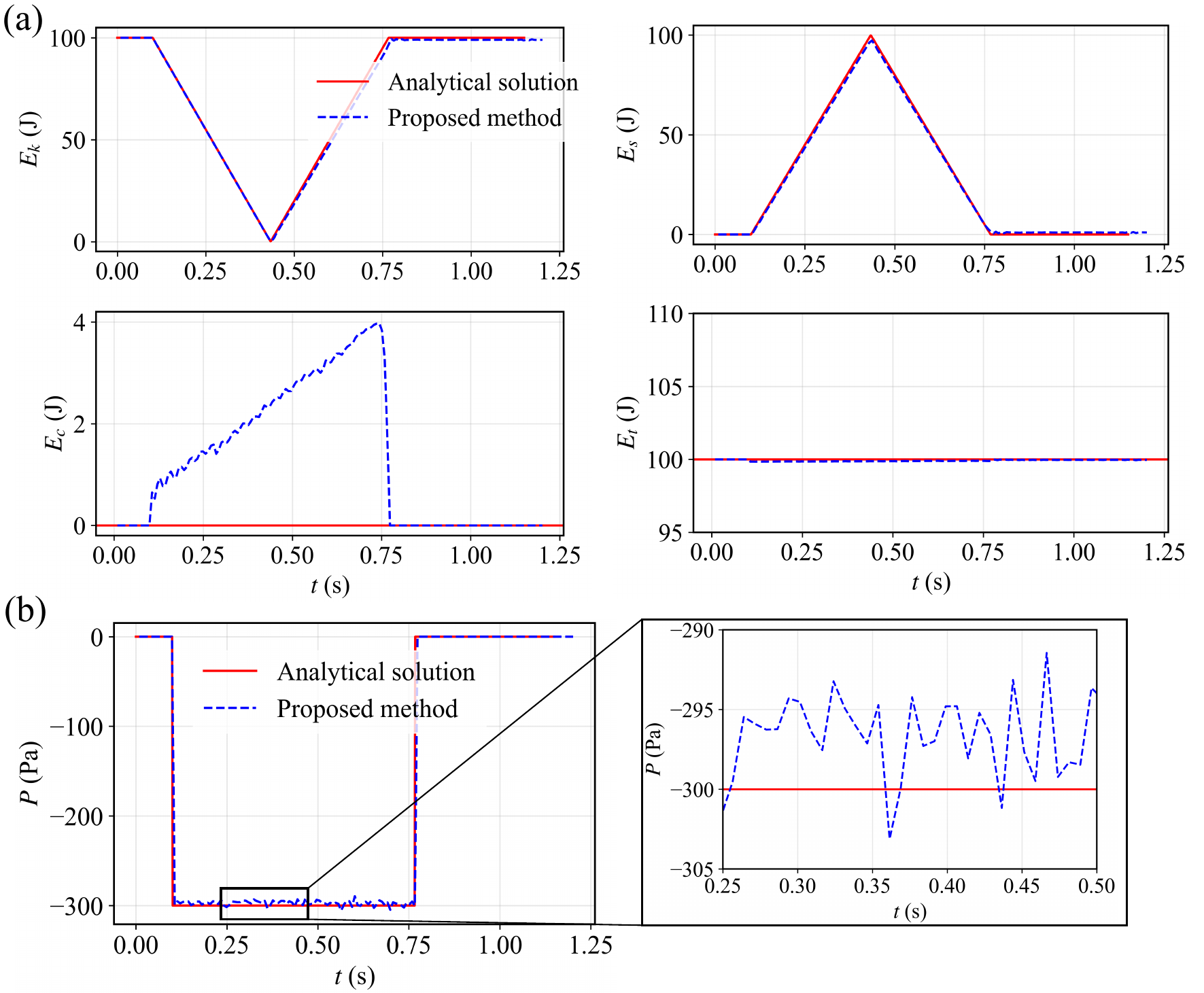}}
    \caption{Impact of an elastic bar: (a) temporal evolution of kinetic energy ($E_k$), strain energy ($E_s$), contact energy ($E_c$), and total energy ($E_t$)  during the entire simulation, (b) distribution of contact pressure at the bar tip.} \label{fig:2.17}
\end{figure}

Figure~\ref{fig:2.17}(a) shows the variation of energy transfer during the impact process. The energies are defined as follows: kinetic energy $E_k = \frac{1}{2}m{v}^2$, strain energy $E_s = \frac{1}{2} \int_V \sigma : \varepsilon dV$, contact energy $E_c = kV_c$ (as defined by our present framework), and total energy $E_t = E_k + E_s + E_c$. Fig.~\ref{fig:2.17}(b) shows the contact pressure distribution at nodes on the contact surface of the bar, with a comparison with the exact solution. We observe that the behavior of our present scheme is in close agreement with the analytical solution in terms of both energy and contact pressure, except that our scheme exhibits small pressure oscillations during the contact phase. This one-dimensional example confirms that the dynamic response is accurately captured and demonstrates the effectiveness of the proposed contact algorithm in handling contact under transient conditions. To further demonstrate the applicability of the proposed approach to dynamic contact problems involving both material and geometric nonlinearities, we next consider two more advanced contact cases. 

\begin{figure}[H] \centering
    {\includegraphics[width=0.85\textwidth]{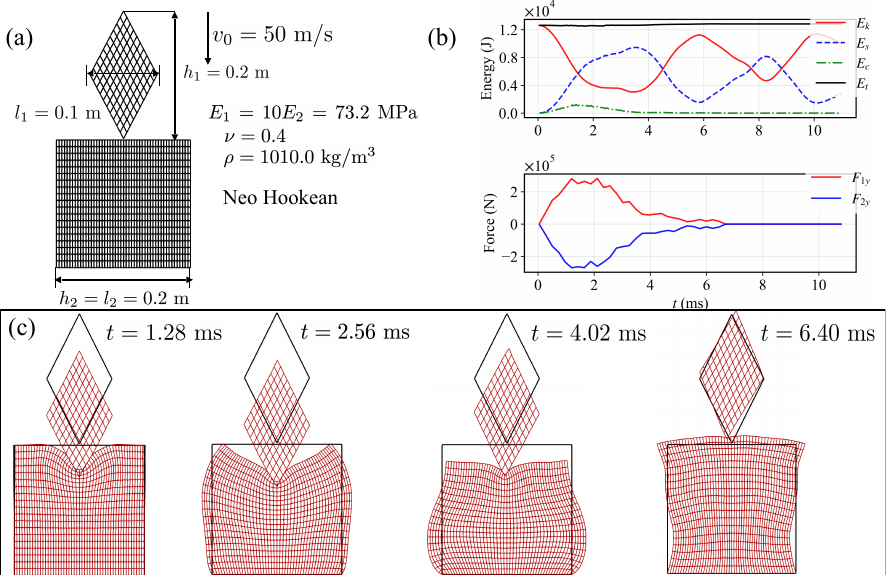}}
    \caption{Collision between an elastic diamond and a block: (a) model setup, (b) temporal evolution of kinetic energy ($E_k$), strain energy ($E_s$), contact energy ($E_c$), and total energy ($E_t$) along with the contact force ($F_{1y}$, $F_{2y}$ ), and (c) snapshots of the deformation at various time instances during the collision.} \label{fig:2.18}
\end{figure}

In the following case, we investigate the finite-deformation response of a hyperelastic diamond colliding with a fixed elastic block, as illustrated in Fig.~\ref{fig:2.18}(a). The bottom surface of the lower block is fully constrained, while the upper diamond is assigned an initial vertical velocity of $v_0 = 50\,\mathrm{m/s}$. Both bodies follow a Neo-Hookean material model with parameters $E_1 = 10 E_2 = 73.2\,\mathrm{GPa}$, $\nu = 0.4$, and $\rho = 1010\,\mathrm{kg/m^3}$. Fig.~\ref{fig:2.18}(b) shows the time evolution of kinetic energy ($E_k$), elastic strain energy ($E_s$), contact energy ($E_c$), and total energy ($E_t$), along with the vertical reaction forces ($F_{1y}$ and $F_{2y}$) on the upper and lower bodies, respectively. The total energy $E_t$ remains within ±1\% of its initial value throughout the simulation, indicating approximate energy conservation. The mutual conversion between $E_k$ and $E_s$ reflects the deformation and rebound of the diamond during the impact. The two bodies make first contact at $t \approx 1\,\mu\mathrm{s}$ and separate at $t \approx 6.78\,\mathrm{ms}$, with the peak contact force reaching approximately $326.3\,\mathrm{kN}$. Snapshots of the deformation configuration at selected time instances are presented in Fig.~\ref{fig:2.18}(c), highlighting key stages of the impact process, including initial contact, maximum compression, and rebound. 

Following the diamond-block impact, we further evaluate the performance of the proposed computational framework in a simulation using a star-shaped indenter, which introduces multiple concave vertices to markedly enhance contact nonlinearities. This configuration evaluates the ability of the proposed framework to handle complex contact interactions under strongly nonlinear conditions, extending the assessment beyond the diamond-block contact scenario.

\begin{figure}[H] \centering
    {\includegraphics[width=0.9\textwidth]{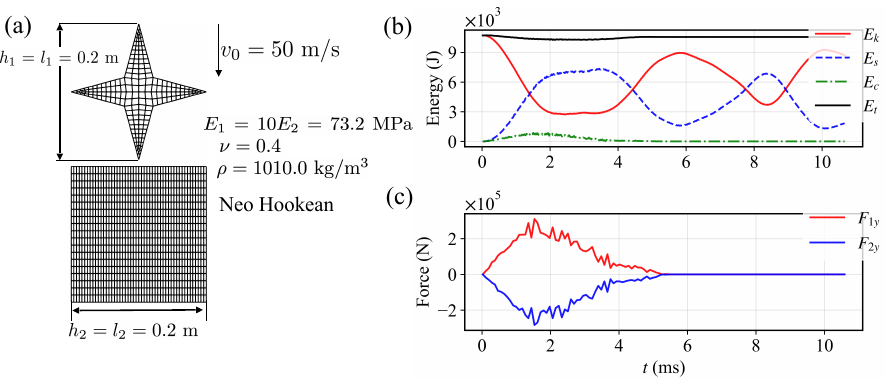}}
    \caption{Collision between an elastic star and a block: (a) model setup, (b) temporal evolution of kinetic energy ($E_k$), strain energy ($E_s$), contact energy ($E_c$), and total energy ($E_t$), and (c) temporal evolution of contact force $F_{1y}$ and $F_{2y}$.} \label{fig:2.19}
\end{figure}

The material properties are set up in the same manner as in the previous example (Fig.~\ref{fig:2.19}(a)). Both the star-shaped indenter and the elastic block follow a Neo-Hookean model with identical parameters. The time evolution of the kinetic energy ($E_k$), elastic energy ($E_s$), contact energy ($E_c$), and total energy ($E_t$) is shown in Fig.~\ref{fig:2.19}(b), while the vertical contact forces are presented in Fig.~\ref{fig:2.19}(c). During contact intervals, a slight energy dissipation is observed, with the total energy $E_t$ decreasing by approximately $0.8\%$ relative to its initial value. This small energy loss arises from numerical discretization errors that are amplified by the non-convex geometry, which can affect the reciprocity of the action–reaction forces (${F}_{1y}$ and ${F}_{2y}$) and consequently introduce a small amount of non-conservative work. In this case, the computed contact force exhibits localized oscillations at the concave vertices due to abrupt changes in the contact region as depicted in Fig.~\ref{fig:2.19}(c); however, the overall dynamic response remains stable, indicating that the framework can handle sharp geometric features without introducing numerical instability.

\begin{figure}[H] \centering
    {\includegraphics[width=1\textwidth]{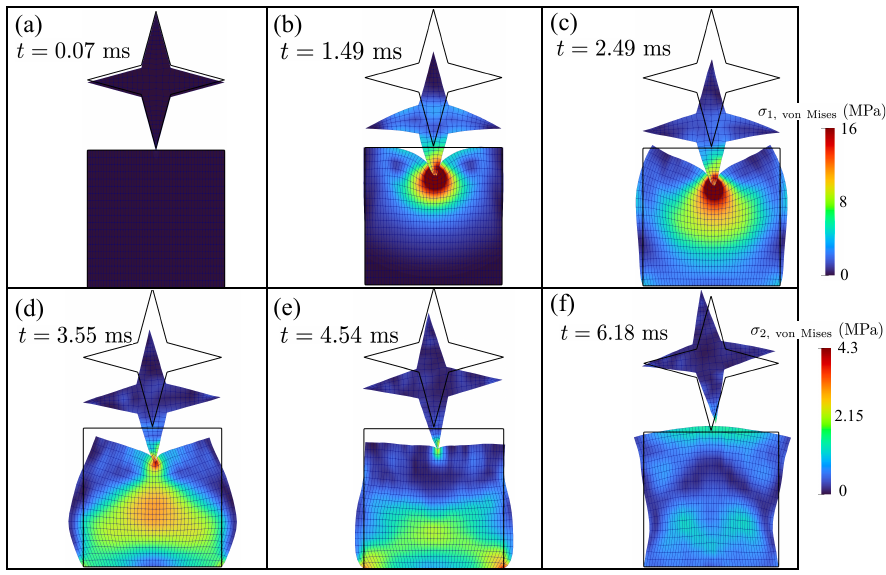}}
    \caption{Snapshots of von Mises stress $\sigma_{\mathrm{vm}}$ at various time instances during the collision.} \label{fig:2.20}
\end{figure}

Snapshots of the von Mises stress $\sigma_{\mathrm{von\ Mises}}$ at selected time instances are provided in Fig.~\ref{fig:2.20}, with subfigures (a)-(f) corresponding to representative stages of the impact and rebound process. The rebound of the star-shaped indenter deviates slightly from the vertical direction, due to minor numerical asymmetries in the computed contact forces arising from both the discretization of the non-convex geometry and the sampling error in Fiber Monte Carlo. Despite the presence of local stress oscillations around the contacting convex tips, the global stress distribution remains smooth and symmetric throughout the contact process. Overall, the results demonstrate that the proposed contact algorithm robustly handles complex, evolving contact configurations in non-convex geometries while ensuring physically accurate energy transfer and numerical stability.

\section{Conclusion}
\label{Sec:conclusion}
This paper presents a finite element contact modeling approach based on the Fiber Monte Carlo method and a volume-based energy-conserving contact model. By defining the contact potential energy as a linear function of the overlap volume between contacting bodies, contact forces are computed directly from the gradient of the potential, without the need to derive a weak form based on contact constraints, and can be directly incorporated into the discretized weak form. In addition, our contact scheme effectively handles complex geometries and operates directly with non-matching meshes, eliminating the need for contact pair searches, master–slave definitions, and projection iterations. The computed results show good agreement with the theoretical Hertz solution, and for contacts involving sharp geometric features (e.g., wedge and conical indenters),  the proposed approach achieves better accuracy in predicting contact pressure and deformation compared to Abaqus. In addition, the framework remains robust when simulating strongly nonlinear contact phenomena, including finite deformations, hyperelastic material behavior, and dynamic collisions.

Despite the favorable performance of our method in handling large-deformation problems with complex geometries, the current implementation does not yet address frictional effects. Future work will focus on extending the framework to include frictional contacts and to support contact-based structural design and optimization. Additionally, a remaining technical challenge is that the current Fiber Monte Carlo-based method does not provide direct access to second-order derivatives, which are necessary for a fully Newton-based solution scheme; addressing this issue is a promising direction for future methodological development.

\section*{Acknowledgements}
\label{sec:acknowledgement}

The authors would like to acknowledge the support from the  Collaborative Research Project Grant (CRPG) by the Research Grants Council of Hong Kong (Project No. C7085-24GF).

\bibliographystyle{elsarticle-num.bst}
\bibliography{bibliography.bib}

\begin{thebibliography}{10}
\expandafter\ifx\csname url\endcsname\relax
  \def\url#1{\texttt{#1}}\fi
\expandafter\ifx\csname urlprefix\endcsname\relax\def\urlprefix{URL }\fi
\expandafter\ifx\csname href\endcsname\relax
  \def\href#1#2{#2} \def\path#1{#1}\fi

\bibitem{wriggers2006computational}
P.~Wriggers, T.~A. Laursen, Computational contact mechanics, Springer, 2006.

\bibitem{wriggers1995finite}
P.~Wriggers, Finite element algorithms for contact problems, Archives of computational methods in engineering 2~(4) (1995) 1--49.

\bibitem{de2017computational}
L.~De~Lorenzis, P.~Wriggers, C.~Wei{\ss}enfels, Computational contact mechanics with the finite element method, Encyclopedia of computational mechanics second edition (2017) 1--45.

\bibitem{shen20222}
W.~Shen, M.~Ohsaki, J.~Zhang, A 2-dimentional contact analysis using second-order virtual element method, Computational Mechanics 70~(2) (2022) 225--245.

\bibitem{sewerin2020finite}
F.~Sewerin, P.~Papadopoulos, On the finite element solution of frictionless contact problems using an exact penalty approach, Computer Methods in Applied Mechanics and Engineering 368 (2020) 113108.

\bibitem{kang2023improved}
S.-H. Kang, S.-M. Lee, S.~Shin, Improved area regularization technique for penalty-method-based node-to-segment contact analysis, Computational Mechanics 71~(4) (2023) 801--825.

\bibitem{peric1992computational}
D.~Peri{\'c}, D.~Owen, Computational model for 3-d contact problems with friction based on the penalty method, International journal for numerical methods in engineering 35~(6) (1992) 1289--1309.

\bibitem{weyler2012contact}
R.~Weyler, J.~Oliver, T.~Sain, J.~Cante, On the contact domain method: A comparison of penalty and lagrange multiplier implementations, Computer methods in applied mechanics and engineering 205 (2012) 68--82.

\bibitem{papadopoulos1998lagrange}
P.~Papadopoulos, J.~Solberg, A lagrange multiplier method for the finite element solution of frictionless contact problems, Mathematical and computer modelling 28~(4-8) (1998) 373--384.

\bibitem{tur2009mortar}
M.~Tur, F.~Fuenmayor, P.~Wriggers, A mortar-based frictional contact formulation for large deformations using lagrange multipliers, Computer Methods in Applied Mechanics and Engineering 198~(37-40) (2009) 2860--2873.

\bibitem{simo1985perturbed}
J.~C. Simo, P.~Wriggers, R.~L. Taylor, A perturbed lagrangian formulation for the finite element solution of contact problems, Computer methods in applied mechanics and engineering 50~(2) (1985) 163--180.

\bibitem{zienkiewicz2005finite}
O.~C. Zienkiewicz, R.~L. Taylor, The finite element method for solid and structural mechanics, Elsevier, 2005.

\bibitem{popp2009finite}
A.~Popp, M.~W. Gee, W.~A. Wall, A finite deformation mortar contact formulation using a primal--dual active set strategy, International Journal for Numerical Methods in Engineering 79~(11) (2009) 1354--1391.

\bibitem{belytschko2014nonlinear}
T.~Belytschko, W.~K. Liu, B.~Moran, K.~Elkhodary, Nonlinear finite elements for continua and structures, John wiley \& sons, 2014.

\bibitem{simo1992augmented}
J.~C. Simo, T.~Laursen, An augmented lagrangian treatment of contact problems involving friction, Computers \& Structures 42~(1) (1992) 97--116.

\bibitem{pietrzak1999large}
G.~Pietrzak, A.~Curnier, Large deformation frictional contact mechanics: continuum formulation and augmented lagrangian treatment, Computer methods in applied mechanics and engineering 177~(3-4) (1999) 351--381.

\bibitem{annavarapu2014nitsche}
C.~Annavarapu, M.~Hautefeuille, J.~E. Dolbow, A nitsche stabilized finite element method for frictional sliding on embedded interfaces. part i: single interface, Computer Methods in Applied Mechanics and Engineering 268 (2014) 417--436.

\bibitem{mlika2017unbiased}
R.~Mlika, Y.~Renard, F.~Chouly, An unbiased nitsche’s formulation of large deformation frictional contact and self-contact, Computer Methods in Applied Mechanics and Engineering 325 (2017) 265--288.

\bibitem{francavilla1975note}
A.~Francavilla, O.~Zienkiewicz, A note on numerical computation of elastic contact problems, International Journal for Numerical Methods in Engineering 9~(4) (1975) 913--924.

\bibitem{hughes1976finite}
T.~J. Hughes, R.~L. Taylor, J.~L. Sackman, A.~Curnier, W.~Kanoknukulchai, A finite element method for a class of contact-impact problems, Computer methods in applied mechanics and engineering 8~(3) (1976) 249--276.

\bibitem{zavarise2009node}
G.~Zavarise, L.~De~Lorenzis, The node-to-segment algorithm for 2d frictionless contact: classical formulation and special cases, Computer Methods in Applied Mechanics and Engineering 198~(41-44) (2009) 3428--3451.

\bibitem{sun2023novel}
C.~Sun, G.~Liu, S.~Huo, G.~Wang, C.~Yu, Z.~Li, A novel node-to-segment algorithm in smoothed finite element method for contact problems, Computational Mechanics 72~(5) (2023) 1029--1057.

\bibitem{HUGHES1976249}
T.~J. Hughes, R.~L. Taylor, J.~L. Sackman, A.~Curnier, W.~Kanoknukulchai, A finite element method for a class of contact-impact problems, Computer Methods in Applied Mechanics and Engineering 8~(3) (1976) 249--276.

\bibitem{puso2020dual}
M.~Puso, J.~Solberg, A dual pass mortar approach for unbiased constraints and self-contact, Computer Methods in Applied Mechanics and Engineering 367 (2020) 113092.

\bibitem{zavarise2009modified}
G.~Zavarise, L.~De~Lorenzis, A modified node-to-segment algorithm passing the contact patch test, International journal for numerical methods in engineering 79~(4) (2009) 379--416.

\bibitem{puso2004mortar}
M.~A. Puso, T.~A. Laursen, A mortar segment-to-segment contact method for large deformation solid mechanics, Computer methods in applied mechanics and engineering 193~(6-8) (2004) 601--629.

\bibitem{neto2014applying}
D.~Neto, M.~Oliveira, L.~Menezes, J.~Alves, Applying nagata patches to smooth discretized surfaces used in 3d frictional contact problems, Computer Methods in Applied Mechanics and Engineering 271 (2014) 296--320.

\bibitem{agrawal2025three}
V.~Agrawal, Three-dimensional varying-order nurbs discretization method for enhanced iga of large deformation frictional contact problems, Computer Methods in Applied Mechanics and Engineering 439 (2025) 117853.

\bibitem{de2011large}
L.~De~Lorenzis, {\.I}.~Temizer, P.~Wriggers, G.~Zavarise, A large deformation frictional contact formulation using nurbs-based isogeometric analysis, International Journal for Numerical Methods in Engineering 87~(13) (2011) 1278--1300.

\bibitem{de2012mortar}
L.~De~Lorenzis, P.~Wriggers, G.~Zavarise, A mortar formulation for 3d large deformation contact using nurbs-based isogeometric analysis and the augmented lagrangian method, Computational Mechanics 49~(1) (2012) 1--20.

\bibitem{aldakheel2020curvilinear}
F.~Aldakheel, B.~Hudobivnik, E.~Artioli, L.~B. da~Veiga, P.~Wriggers, Curvilinear virtual elements for contact mechanics, Computer Methods in Applied Mechanics and Engineering 372 (2020) 113394.

\bibitem{wriggers2019virtual}
P.~Wriggers, W.~T. Rust, A virtual element method for frictional contact including large deformations, Engineering Computations 36~(7) (2019) 2133--2161.

\bibitem{zavarise1998segment}
G.~Zavarise, P.~Wriggers, A segment-to-segment contact strategy, Mathematical and Computer Modelling 28~(4-8) (1998) 497--515.

\bibitem{el2001stability}
N.~El-Abbasi, K.-J. Bathe, Stability and patch test performance of contact discretizations and a new solution algorithm, Computers \& Structures 79~(16) (2001) 1473--1486.

\bibitem{bathe2001inf}
K.-J. Bathe, The inf--sup condition and its evaluation for mixed finite element methods, Computers \& structures 79~(2) (2001) 243--252.

\bibitem{feng2021energy}
Y.~Feng, An energy-conserving contact theory for discrete element modelling of arbitrarily shaped particles: Basic framework and general contact model, Computer Methods in Applied Mechanics and Engineering 373 (2021) 113454.

\bibitem{richardson2024fiber}
N.~Richardson, D.~Oktay, Y.~Ovadia, J.~C. Bowden, R.~P. Adams, Fiber monte carlo, in: The Twelfth International Conference on Learning Representations, 2024.

\bibitem{izmailov2003karush}
A.~F. Izmailov, M.~V. Solodov, Karush-kuhn-tucker systems: regularity conditions, error bounds and a class of newton-type methods, Mathematical Programming 95~(3) (2003) 631--650.

\bibitem{marsden1994mathematical}
J.~E. Marsden, T.~J. Hughes, Mathematical foundations of elasticity, Courier Corporation, 1994.

\bibitem{FENG2021113454}
Y.~Feng, An energy-conserving contact theory for discrete element modelling of arbitrarily shaped particles: Basic framework and general contact model, Computer Methods in Applied Mechanics and Engineering 373 (2021) 113454.

\bibitem{mohamed2020monte}
S.~Mohamed, M.~Rosca, M.~Figurnov, A.~Mnih, Monte carlo gradient estimation in machine learning, Journal of Machine Learning Research 21~(132) (2020) 1--62.

\bibitem{mooney1997monte}
C.~Z. Mooney, Monte carlo simulation, Sage, 1997.

\bibitem{xue2023jax}
T.~Xue, S.~Liao, Z.~Gan, C.~Park, X.~Xie, W.~K. Liu, J.~Cao, Jax-fem: A differentiable gpu-accelerated 3d finite element solver for automatic inverse design and mechanistic data science, Computer Physics Communications (2023) 108802.

\bibitem{haslinger1981contact}
J.~Haslinger, I.~Hlav{\'a}{\v{c}}ek, Contact between elastic bodies. ii. finite element analysis, Aplikace Matematiky 26~(4) (1981) 263--290.

\bibitem{taylor2013finite}
R.~L. Taylor, O.~C. Zienkiewicz, The finite element method, Butterworth-Heinemann Oxford, UK:, 2013.

\bibitem{hairer2006structure}
E.~Hairer, C.~Lubich, G.~Wanner, Structure-preserving algorithms for ordinary differential equations, Geometric numerical integration 31.

\bibitem{laursen2003computational}
T.~A. Laursen, Computational contact and impact mechanics: fundamentals of modeling interfacial phenomena in nonlinear finite element analysis, Springer Science \& Business Media, 2003.

\bibitem{liang2024mortar}
W.~Liang, H.~Fang, Z.-Y. Yin, J.~Zhao, A mortar segment-to-segment frictional contact approach in material point method, Computer Methods in Applied Mechanics and Engineering 431 (2024) 117294.

\bibitem{johnson1987contact}
K.~L. Johnson, Contact Mechanics, Cambridge University Press, 1985.

\bibitem{popov2025handbook}
V.~L. Popov, M.~He{\ss}, E.~Willert, Handbook of Plane Contact Mechanics, Springer, 2025.

\bibitem{truman1995contact}
C.~Truman, A.~Sackfield, D.~Hills, Contact mechanics of wedge and cone indenters, International Journal of Mechanical Sciences 37~(3) (1995) 261--275.

\bibitem{popov2019handbook}
V.~L. Popov, M.~He{\ss}, E.~Willert, Handbook of contact mechanics: exact solutions of axisymmetric contact problems, Springer Nature, 2019.

\bibitem{jin2016node}
S.~Jin, D.~Sohn, S.~Im, Node-to-node scheme for three-dimensional contact mechanics using polyhedral type variable-node elements, Computer Methods in Applied Mechanics and Engineering 304 (2016) 217--242.

\end{thebibliography}

\begin{appendices}

\renewcommand{\thefigure}{\thesection.\arabic{figure}}
\renewcommand{\thetable}{\thesection.\arabic{table}}
\renewcommand{\theequation}{\thesection.\arabic{equation}}

\setcounter{figure}{0}  
\setcounter{table}{0}   
\setcounter{equation}{0} 
\section{Sampling formulation}
\label{App:sample_method}

Given a sampling domain $\Omega \subset \mathbb{R}^d$, each fiber is defined by a starting point $\boldsymbol{x_s}$ sampled uniformly over $\Omega$ and an endpoint on the surface of a norm ball of radius $\ell>0$ centered at $\boldsymbol{x_s}$. To ensure that all endpoints lie within a valid sampling region, we introduce the extended domain $\bar{\Omega} = \{\boldsymbol{x} \mid \mathrm{dist}(\boldsymbol{x}, \Omega) \le \ell \}$, which adds a shell of width $\ell$ around $\Omega$ and preserves the uniformity of the fiber distribution. The sampling process can be expressed as:

\begin{equation}
\boldsymbol{x}_s \sim \text{Uniform}(\Omega), \quad \boldsymbol{q} \sim \mathrm{N}(0, \mathbb{I}_d)
\end{equation}
where the endpoint $\boldsymbol{x_e} \in \bar{\Omega}$ of the fiber is determined as:

\begin{equation}
\boldsymbol{x}_e = \boldsymbol{x}_s + \ell \frac{\boldsymbol{q}}{\|\boldsymbol{q}\|}
\label{eq:fiber_endpoint}
\end{equation}
The expectation estimated by FMC is formed as the sample expectation of its line integral over a collection of $n$ fibers of length $\ell$:

\begin{equation}
\mathbb{E}_{\boldsymbol{x}\sim\Omega}[h_\theta(\boldsymbol{x})] = \frac{1}{n} \sum_{i=1}^{n} \int_{\boldsymbol{x}_s}^{\boldsymbol{x}_e} h(\boldsymbol{x}) \, p(\boldsymbol{x}) \, ds 
= \frac{1}{n} \sum_{i=1}^{n} \int_0^1 h(\boldsymbol{r_i}(t))\, |\boldsymbol{r_i}'(t)| / \ell \, dt
\end{equation}
Here, $\boldsymbol{r_i}(t)$ is a bijective parameterization of the $i$-th fiber, defined by $\boldsymbol{r_i}(t) = \boldsymbol{x_s} + t(\boldsymbol{x_e} - \boldsymbol{x_s})$ for $t \in [0,1]$. $h_\theta(\boldsymbol{x})$ is the function to be integrated. The marginal density along a fiber, $p(\boldsymbol{x})$, is induced by sampling over the extended domain $\bar{\Omega}$, and is given by $p(\boldsymbol{x}) = 1/\ell$, which ensures uniform sampling along each fiber.

\setcounter{figure}{0}  
\setcounter{table}{0}  
\setcounter{equation}{0} 
\section{Implicit function formulation}
\label{App:Implicit function formulation}
Implicit functions provide a flexible parameterization for complex contact boundaries. The domain of a contact body is represented using the zero sublevel set of a scalar field $g: \mathbb{R}^d \to \mathbb{R}$ with parameters $\theta \in \mathbb{R}^m$, which typically represents the displacement field in our contact problem. This zero sublevel set is defined as
\begin{equation}
g_z \equiv \{\boldsymbol{x} \in \mathbb{R}^d \mid g_\theta(\boldsymbol{x}) \leq 0\}
\end{equation}
with its boundary (denoted $\textbf{bd}\, g_z$) explicitly defining the contact boundary:
\begin{equation}
\mathbf{bd}\, g_z \equiv \{ \boldsymbol{x} \in \mathbb{R}^d \mid g_\theta(\boldsymbol{x}) = 0 \}. 
\end{equation}
A fiber that intersects the contact boundary $\mathbf{bd}\, g_z$ is considered. We define the interpolant $h: \mathbb{R}^m \to (0,1)$ as a function of the parameters $\theta$, such that the convex combination
\begin{equation}\label{eq:interpolant}
\boldsymbol{\alpha} = \boldsymbol{x_s} + h(\theta)(\boldsymbol{x_e} - \boldsymbol{x_s})
\end{equation}
lies on the fiber and satisfies $g_\theta(\boldsymbol{\alpha}) = 0$, so that $g_\theta(\boldsymbol{\alpha}) \in \mathbf{bd}\, g_z$. The corresponding optimality condition is then defined by
\begin{equation}\label{eq:constraints}
L(h(\theta), \theta) = g_\theta(\boldsymbol{\alpha}) = 0,
\end{equation}
which implicitly determines the intersection point along the fiber.

Consider an objective function $\mathcal{J}: R^m \to R$ that depends on $\theta$ through a variable $h(\theta)$. Its total derivative with respect to $\theta$ follows from the chain rule:
\begin{equation}\label{eq:chain rule}
\frac{d\mathcal{J}}{d\theta} = \frac{\partial \mathcal{J}}{\partial h}\frac{dh}{d\theta} + \frac{\partial \mathcal{J}}{\partial \theta}.
\end{equation}
If $h(\theta)$ satisfies the optimality condition $L(h(\theta),\theta) = 0$, the derivative $\frac{dh}{d\theta}$ can be obtained by implicit differentiation:
\begin{equation}\label{eq:implicit}
\frac{\partial L}{\partial h}\frac{dh}{d\theta} = -\frac{\partial L}{\partial \theta}.
\end{equation}
A fully differentiable framework for computing the points of intersection is equipped. 

\end{appendices}

\end{document}